\documentclass[aps,footinbib,superscriptaddress]{revtex4-2}
\usepackage[margin=1in]{geometry}
\usepackage{amsmath,amsfonts, setspace,graphicx,comment,chemformula}
\usepackage{array}
\usepackage{soul,xcolor,algpseudocode,algorithm}
\usepackage{subfig,mwe}
\usepackage{multirow,hhline}
\usepackage[section]{placeins}
\DeclareMathOperator*{\argmax}{arg\,max}
\DeclareMathOperator*{\argmin}{arg\,min}
\newcommand{\be}{\begin{equation}}
\newcommand{\ee}{\end{equation}}

\makeatletter
\AtBeginDocument{%
  \expandafter\renewcommand\expandafter\subsection\expandafter{%
    \expandafter\@fb@secFB\subsection
  }%
}
\makeatother

\begin{document}

\title{Optimal policies for Bayesian olfactory search in turbulent flows}
\author{R.\ A.\ Heinonen}
\affiliation{Dept.\ Physics and INFN, University of Rome ``Tor Vergata'', Via della Ricerca Scientifica 1, 00133 Rome, Italy}
\author{L.\ Biferale}
\affiliation{Dept.\ Physics and INFN, University of Rome ``Tor Vergata'', Via della Ricerca Scientifica 1, 00133 Rome, Italy}
\author{A.\ Celani}
\affiliation{Quantitative Life Sciences, The Abdus Salam International Centre for Theoretical Physics, 34151 Trieste, Italy}
\author{M. Vergassola}
\affiliation{Laboratoire de physique, \'Ecole Normale Sup\'erieure, CNRS, PSL Research University, Sorbonne University, Paris 75005, France}
\begin{abstract}
In many practical scenarios, a flying insect must search for the source of an emitted cue which is advected by the atmospheric wind.  On the macroscopic scales of interest, turbulence tends to mix the cue into patches of relatively high concentration over a background of very low concentration, so that the insect will only detect the cue intermittently and cannot rely on chemotactic strategies which simply climb the concentration gradient. In this work, we cast this search problem in the language of a partially observable Markov decision process (POMDP) and use the Perseus algorithm to compute strategies that are near-optimal with respect to the arrival time. We test the computed strategies on a large two-dimensional grid, present the resulting trajectories and arrival time statistics, and compare these to the corresponding results for several heuristic strategies, including (space-aware) infotaxis, Thompson sampling, and QMDP. We find that the near-optimal policy found by our implementation of Perseus outperforms all heuristics we test by several measures. We use the near-optimal policy to study how the search difficulty depends on the starting location. We discuss additionally the choice of initial belief and the robustness of the policies to changes in the environment. Finally, we present a detailed and pedagogical discussion about the implementation of the Perseus algorithm, including the benefits --- and pitfalls --- of employing a reward shaping function.
\end{abstract}
\maketitle

\section{Introduction}
Certain flying insects depend on a remarkable ability to use olfactory cues to locate distant sources. Two salient, well-studied examples are female mosquitoes, which use a combination of carbon dioxide and other cues such as odors and heat \cite{mcmeniman2014,carde2015} to find their human hosts, and moths, which track potential mates using emitted sex pheromones to which they are extremely sensitive. 

In experiments, mosquitoes immediately begin flying upwind in the presence of fluctuating CO$_{2}$ plumes \cite{healy1995,geier1999,dekker2011} from a distance of up to tens of meters away from the source \cite{gillies1980}, from which distances visual cues are not useful \cite{bidlingmayer1980,vanbreugel2015}. Male moths exhibit similar upwind search behavior when exposed to pheromones from females \cite{carde1979,kennedy1981}, and their  maximum effective search range is even further, on the order of 100 m \cite{elkinton1987}. Thus the relevant lengthscales for the search are macroscopic. On such lengthscales, effects due to turbulence dominate the transport of passive scalars; the turbulent transport induces a wildly fluctuating, intermittent concentration landscape. A fixed location tens of meters from a source may go long times --- tens of seconds --- without a detectable increase in local concentration \cite{yee1993}. Therefore, the insect cannot quickly estimate the concentration gradient, and simple search strategies like chemotaxis which rely on moving up the local gradient have minimal efficacy. A searching insect must then make effective use of the limited information it can glean from intermittent detections \cite{reddy2022}. Information-theoretic studies suggest that the insect is best served by measuring concentrations as coarsely as possible, i.e.\ a binary low/high signal relative to some threshold \cite{boie2018,victor2019}.

In addition to insects, many mammals, such as dogs and rodents, alternate between sniffing the ground and sniffing the air when performing olfactory navigation \cite{thesen1993,khan2012,gire2016,jinn2020}. The latter behavior implies that the mammals integrate airborne cues into their search and that they may depend in part on the same fundamental strategies as insects.

Both moths and mosquitoes exhibit similar behavior during their upwind flight, including a tendency towards zigzagging motion across the mean wind \cite{kennedy1983}, or ``casting''; fruit flies also exhibit similar behavior in response to attractive odor stimuli \cite{budick2006,vanbreugel2014}). This suggests that good strategies for olfactory search are universal and above all depend on the physics underlying the dispersal of the attractant cues. Inspired by this observation, Balkovsky and Shraiman \cite{balkovsky2002} proposed \emph{cast-and-surge}, a simple heuristic policy for olfactory search. Cast-and-surge combines cross-wind casting, which helps to locate the downwind axis of the source where detections are most probable, with ``surging,'' direct upwind motion immediately after a detection. 

The cast-and-surge decision-making algorithm depends only on the detection history. Another class of policies are \emph{model-based}, in the sense that they take as input some hard-wired model for the statistics of detection, which one can imagine may be instinctual to an insect. An important example is \emph{infotaxis} \cite{infotaxis}, which chooses actions by maximizing the information about the source location which, based on the information the agent currently has, is expected to be gained. Trajectories generated using infotaxis encompass a number of behaviors, including casting.

While a zoology of model-based heuristics exists, when one has an exact or approximate expression for the detection probability, a natural question to ask is what kind of policy is \emph{optimal} --- that is, results in a minimum mean arrival time --- given the statistics? What behaviors are seen in this optimal policy? And how do the various heuristic policies compare? 

The mathematical language of partially observable Markov decision processes (POMDPs) allows us to formalize the search problem and specify the optimal strategy as the solution to the Bellman equation, a nonlinear functional equation. However, solving for the optimal policy in a POMDP is known to be a difficult computational challenge, and scalability to large problem sizes is a key issue. Recently, a promising effort \cite{loisy2021} to solve the problem using a variant of deep reinforcement learning to obtain an approximate solution of the Bellman equation; however, it was left uncertain how effectively this approach scaled to large problem sizes, and the authors did not include a mean wind.

In this work, we use a POMDP algorithm called Perseus \cite{perseus}, coupled with reward potential shaping \cite{ng1999}, to approximate the solution to the Bellman equation and obtain near-optimal policies for olfactory search on a large grid with thousands of points. Perseus was previously used for the search problem in Ref.~\cite{reddy2022}, but the results presented were limited and purely qualitative. Presently, we build model environments with three different characteristic emission rates, compute near-optimal policies for each environments, and compare the arrival-time statistics of the near-optimal policies with those of a few interesting heuristic policies (including two versions of infotaxis). We find that the near-optimal policy found by our implementation of Perseus successfully outperforms all the tested heuristics in each environment. This establishes a reasonably scalable baseline for the POMDP solution of this problem, allows for the study of the behaviors and statistics which characterize optimal search, and opens the possibility to apply the same algorithm to more complex search problems, for example, the case of multiple sources. We also suggest that the present work may have applications to robotics for the purpose of detection of hazardous chemicals or explosives; previous studies saw the design of robots for olfactory search using reactive policies \cite{kuwana1999,pyk2006} and using infotaxis \cite{masson2013,martinez2013}.

The paper is organized as follows. In Sec.~\ref{sec:problem}, we detail the search problem and its assumptions, and we introduce the simple mathematical model for detections which we use in this work. In Sec.~\ref{sec:methods}, we review the POMDP formalism, cast the search problem in this language, and describe both the heuristic policies and the methods used to solve the POMDP directly. Important details include the choice of initial condition and the use of a reward shaping function to accelerate convergence. In Sec.~\ref{sec:results}, we present example trajectories and arrival time statistics for the various policies, including mean arrival times for several test problems and detailed pdfs. We also test the robustness of the policies under changes in the model environment and use the near-optimal policies to obtain the approximate best mean arrival time as a function of starting position. Finally, in Sec.~\ref{sec:discussion} we summarize our results and discuss avenues for future research.

\section{Problem description}\label{sec:problem}
We will consider the problem of a model insect, or agent, searching for a stationary source located somewhere upwind, using some olfactory cue. We will constrain the motion of the agent to a two-dimensional plane, and we will discretize space into a rectangular grid (see Fig. \ref{fig:schematic}). While this constraint is primarily made for computational ease (since very large POMDPs are extremely difficult to solve), it also models the fact that for problems of interest the source will typically be close to the ground. We will assume the agent begins its search after a detection event (see Sec.~\ref{sec:init} for more details on the initialization) and stops when it reaches the gridpoint corresponding to the source location. The goal is to find a search strategy which minimizes the mean arrival time to the source.

At each timestep, the agent will make either a detection or a nondetection of the cue, which may be interpreted as the insect observing a concentration above threshold. Due to the random nature of turbulent mixing, detection will occur with some space-dependent probability, which will be small far enough downwind from the source and nearly vanishing upwind of the source \cite{celani2014}.We assume the Markov property: the detection events are independent in time and space. In principle, we could allow for multiple detections per timestep, but in the most interesting case, detections are rare enough that more than one detection in a small time interval is exceedingly unlikely (unless the agent is very close to the source).

Physically, the discretization may be understood as assuming the agent flies at a fixed speed $v$ and measures the local concentration by integrating over a characteristic sampling time $\Delta t$ (the decision timestep). The gridspacing is then $\Delta x = v\Delta t.$ 

We seek an expression for the mean rate of detections in a statistically-steady turbulent flow, as a function of spatial position. We model the turbulent environment with an effective turbulent diffusivity $D$ (assumed to be much larger than the collisional viscosity). We impose a mean wind $V\hat{x}$ and fix a source with emission rate $S$ at the origin, $\mathbf{r}_0=0$. The advection-diffusion equation is then
\begin{equation}
\label{eq:diffusion}
\partial_t c + V \partial_x c = D \nabla^2 c + S \delta(\mathbf{r}) - c/\tau,
\end{equation}
where $c$ is the concentration field and $\tau$ is a particle lifetime that can be identified as a turbulent mixing time or coherence time. A simple dimensional estimate for the turbulent diffusivity is $D \sim \ell  \tilde v$, where $\tilde v$ is the rms fluctuation velocity and $\ell$ is the turbulent mixing length. In this framework, we then have $\tau \sim \ell/\tilde v$.
In three spatial dimensions, the stationary solution of Eq.~\ref{eq:diffusion} is 
\begin{equation}
\label{eq:concentration}
c(\mathbf{r})= \frac{S}{4 \pi D |\mathbf{r} | } \exp\left( \frac{V \mathbf{r} \cdot \hat x}{2 D} - \frac{|\mathbf{r}|}{\lambda}\right)
\end{equation}
with $\lambda \equiv \sqrt{D \tau/(1+ V^2\tau/4D)}$.
Using Smoluchowski's expression for the rate of encounters of a sphere of radius $a$ with molecules diffusing with diffusivity $D$ \cite{smoluchowski1918}, we obtain an estimated number of encounters in a time interval $\Delta t$
\begin{equation}
h(\mathbf{r}) =  4\pi a D\Delta t \, c(\mathbf{r}) 
\label{eq:encounters}
\end{equation}
where $a$ is a characteristic lengthscale of the searcher. We then imagine that the animal, which is constrained to the plane of the source, treats the number of detections per timestep as a Poisson variable with rate $h$.

In this work, the agent will search in a toy environment wherein the detections are generated artificially, drawn from the distribution specified by the diffusive model. (Understanding the robustness of policies trained using the present model when applied to more realistic environments remains one open important problem that goes well beyond the scope of this paper.) But of course, the model is a simplification of the turbulence physics; see, for example, \cite{celani2014} for a more sophisticated treatment. In reality, the dynamics of the odor molecules will be neither purely diffusive nor purely ballistic, and moreover detection events will have a nonzero spatiotemporal correlation. On the other hand, the present diffusive model has seen significant use in past work (e.g.\ \cite{infotaxis,masson2009,loisy2021}) and in any case leads to a good benchmark search problem which is far from trivial to solve. Above all, the model captures the key phenomenological features that make the problem difficult and interesting \cite{celani2014}: detections are \emph{stochastic} (with some space-dependent probability) and \emph{rare} enough that fine-grained information about the local concentration field is not very useful. As long as correlations are neglected, differences between the toy model and real data will be quantitative rather than qualitative.

After one introduces a grid-spacing $\Delta x$, the model can be parametrized by three nondimensional quantities: the nondimensional emission rate $\bar S \equiv a S \Delta t/ \Delta x$, the nondimensional mean wind $V\Delta x/D$, and the nondimensional coherence time $V^2 \tau/D$. The values we use for these quantities are shown in Table~\ref{table:parameters}.

A final, important assumption is that the agent has knowledge (say, through instinct) of the detection statistics implied by Eqs.~\ref{eq:concentration}--\ref{eq:encounters}. This will be necessary to perform Bayesian inference. 

\begin{table}
\begin{tabular}{|c|c|c|c|}
\hline
category & (hyper)parameter & description & value(s) used\\
\hline
\multirow{6}{*}{Environment} & $V\Delta x/D$ & nondimensional mean wind & 2 \\
& $V^2 \tau/D$& nondimensional turb.\ coherence time & 150 \\
& $\bar{S} \equiv a \Delta t S/\Delta x$ & nondimensional emission rate & 0.25, 2.5, 25 \\
& $\Delta x/\Delta y$ & grid-spacing ratio & 1 \\
& $N_x$ & grid points along wind axis & 81 \\
& $N_y $& grid points along cross-wind axis & 41 \\
& $\mathbf{r}_0 $ & source position & $(10,20)$ (grid spacing units) \\ 
\hline
\multirow{5}{*}{POMDP solution} & $|{\cal B}|$ & number of Perseus beliefs & 45000 \\
& $g(D)$ & reward shaping function & $0.001 D^2, 0.1 D$\\
& $\pi_{B}$ & policy for belief collection & infotaxis \\
& $T_{\rm wait}$ & max time to wait for first detection & 1000 \\
& $\gamma$ & discount factor & 0.96, 0.98 \\
\hline

\end{tabular}
\caption{Table of parameters and hyperparameters, relating to the turbulence physics, the POMDP definition and grid, and the Perseus algorithm.}\label{table:parameters}
\end{table}

\section{Methods}\label{sec:methods}

\subsection{POMDP setup}
We now cast the search problem in the language of a partially observable Markov decision process (POMDP) \cite{astrom1965,kaelbling1998}. The fundamental ingredients of a POMDP are a state space $S$, a set of actions $A$, a set of observations $O$, and a reward function
\[ R: S\times A \to \mathbb{R}. \]
At each timestep, the \emph{agent} is in some state $s\in S$ and selects an action $a\in A,$ which causes the agent to transition from state $s$ to $s'$ with a specified probability ${\rm Pr}(s'|s,a).$ 

\begin{figure}
    \centering
    \includegraphics[width=\linewidth]{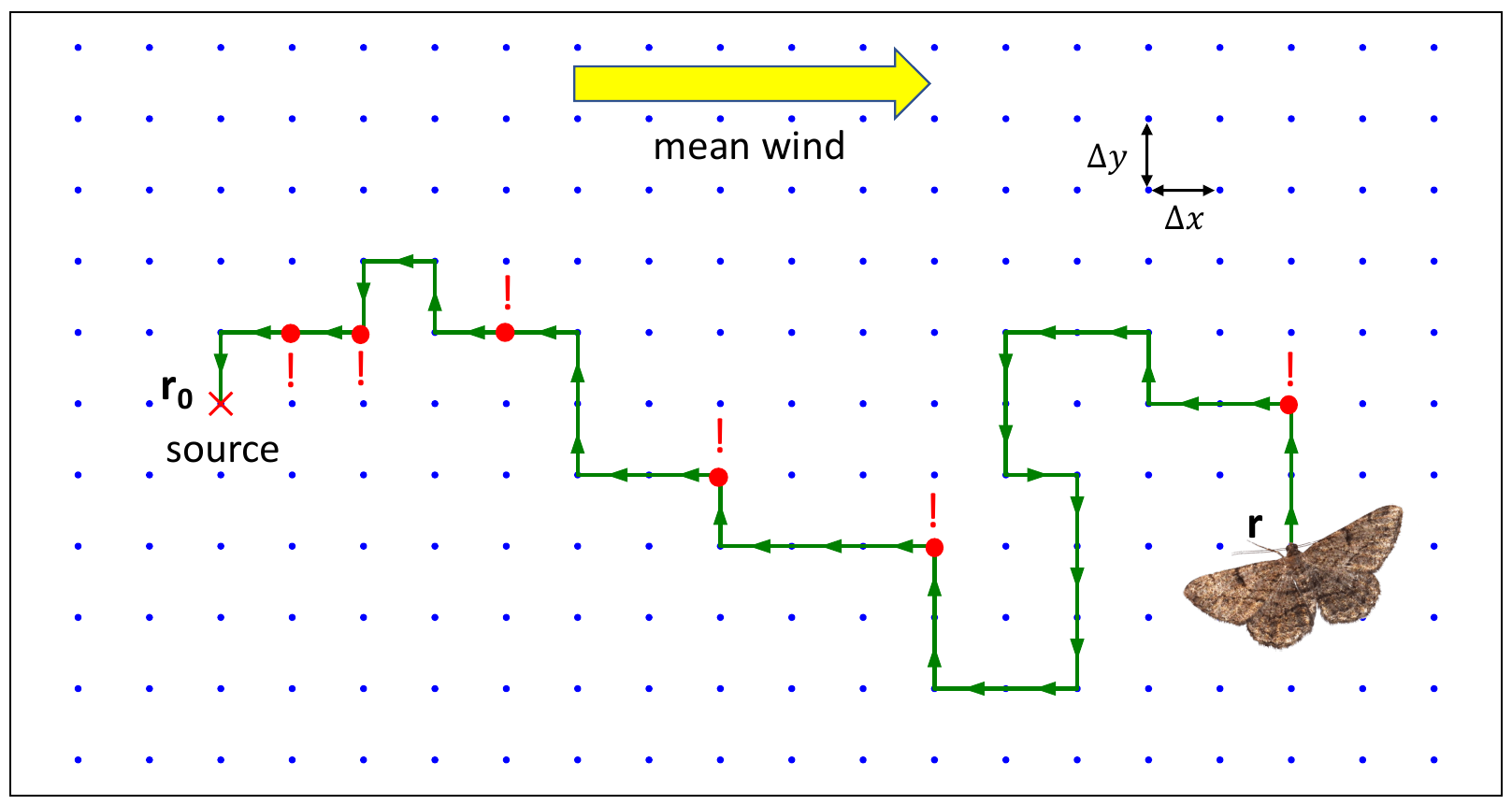}
    \caption{Basic schematic of the search POMDP. The model insect searches for the source (red X) on a 2-D grid, in the presence of a mean wind. At each time step, it either detects the cue (red circle with ``!'') or makes no detection, with a specified probability; these observations are used to update the belief of where the source is. It then moves to an adjacent gridpoint. The search terminates when it finds the source, or when some maximum search time is exceeded.}
    \label{fig:schematic}
\end{figure}

For our purposes, the agent is a model insect living on an $N_x \times N_y$ grid with spacings $\Delta x$ and $\Delta y$ (see Fig.~\ref{fig:schematic}). The source is fixed at a point $\mathbf{r}_0$. The state of the agent is its relative position with respect to the source $\mathbf{s} \equiv \mathbf{r}-\mathbf{r}_0$, with $\mathbf{r}$ the agent location. The actions are to simply to move to an adjacent grid point, $A= \{ (\Delta x ,0), (-\Delta x,0), (0,\Delta y), (0,-\Delta y)\},$ so that after taking action $\mathbf{a}$ the new state is $\mathbf{s}' = \mathbf{s} + \mathbf{a}$ with probability 1 --- unless it attempts to leave the grid, in which case the agent is unmoved, or if it has found the source. The state of occupying the same gridpoint as the source is treated as an \emph{absorbing} state, which is to say that no action will change the state. (more details are given in the Appendix.)

The agent also receives a reward $R(s,a)$ for taking an action, which is \emph{discounted} by a factor $0<\gamma < 1$. We set the reward to be unity for finding the source, and zero otherwise (other choices are easily seen to be equivalent, provided $\gamma<1$ --- see Appendix \ref{app:rewards}). This sets up an optimization problem, namely to craft a \emph{policy} for choosing actions which maximizes the expected total reward
\be
\mathbb{E} [R_{tot}] = \sum_{t=0}^\infty \gamma^t \mathbb{E}[R(s_t,a_t)] =  \mathbb{E}[\gamma^{T-1}],
\ee
where $T$ is the arrival time to the source. The discount factor helps to regularize POMDPs by reducing the influence of times far in the future, and its value sets the extent to which the agent should prioritize immediate rewards vis-\`a-vis future rewards. This preference for short- or long-term rewards is quantified by a characteristic time called the \emph{horizon}  $\sim 1/\log \gamma^{-1} \simeq 1/(1-\gamma)$. Rewards which are earned at times in the future beyond the horizon are suppressed in the decision-making process. We refer the reader to Appendix \ref{app:rewards} for further discussion.

So far, we have only described a basic Markov decision process (MDP). The challenge of partial observability is that the agent does not have access to its state, and instead maintains a probability distribution over $S$ called a \emph{belief}, $b(s)$. This is of course relevant to the present search problem because the agent does not know where the source is. The belief lives in a $|S|-1$ dimensional simplex: the set of vectors in $S$ with nonnegative components summing to one. At each timestep, after taking its action, the agent makes an observation $o\in O$ 
which is used to update the belief using Bayes' rule
\be
\label{eq:bayes}
b_{o,a}(s')= \frac{{\rm Pr} (o|s',a) \sum_s b(s) {\rm Pr} (s'|s,a)}{\sum_{s,s'} {\rm Pr}(o|s',a)  b(s) {\rm Pr} (s'|s,a)},
\ee
where $b_{o,a}$ denotes the updated belief after taking action $a$ and making observation $o$. 
Note that in our problem, the transition probability is deterministic and, excluding the aforementioned edge cases where the agent has already found the source or tries to exit the grid, we can write $\mathrm{Pr}(s'|s,a)= \delta_{s',s+a}$. Also, the observation likelihood is independent of the previous action taken by the agent and depends only on its relative displacement from the origin, $\mathrm{Pr}(o|s,a)=\mathrm{Pr}(o|s)$. However, we will usually give expressions pertaining to POMDPs in a general form when possible. 

\begin{figure}
    \centering
    \includegraphics[width=\linewidth]{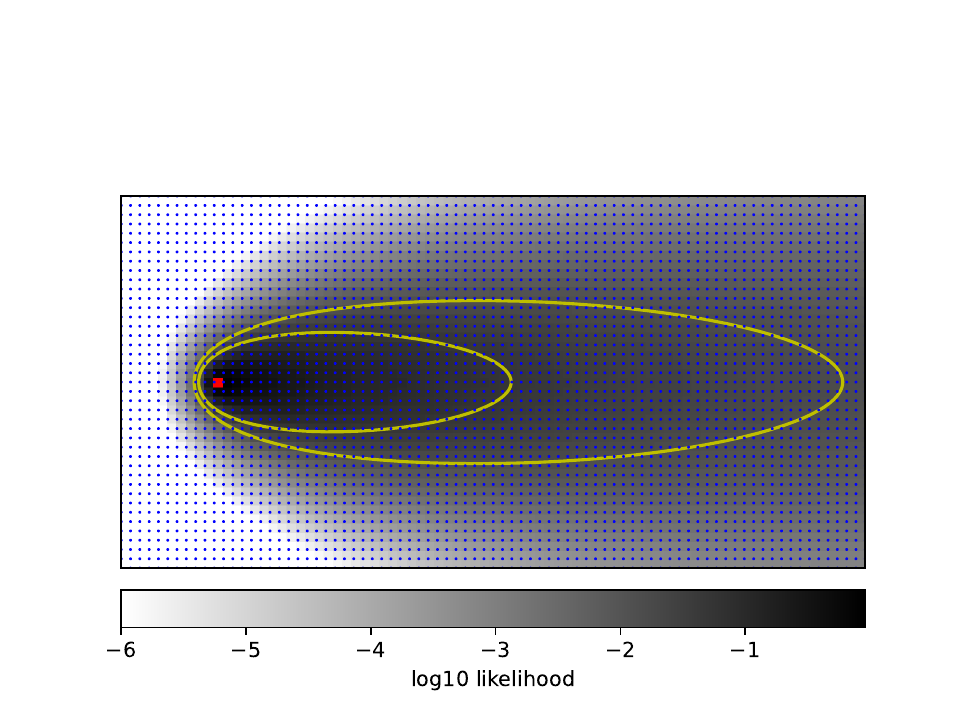}
    \caption{Plot showing the gridworld (blue dots) overlaid with the log detection likelihood for our choice of model parameters when $\bar{S} = 2.5$. The source, in red, has zero detection likelihood because it triggers a special observation. The agent always starts its search within two selected likelihood isocurves, shown in yellow.}
    \label{fig:ellipses}
\end{figure}

The likelihood is set by the diffusive model of Sec.~\ref{sec:problem} as follows. We define three possible observations. First, the agent may discover that it has found the source, which occurs with probability $p_s=\delta_{\mathbf{s},0}$, where $\delta$ is a Kronecker function. Otherwise, the agent may observe either a detection with probability
\begin{equation}
\label{eq:det}
\mathrm{Pr}(o={\rm det.} | \mathbf{s})= (1- \exp(-h(\mathbf{s}))) (1- p_s)
\end{equation} 
or a nondetection with probability
\begin{equation}
\label{eq:nondet}
\mathrm{Pr}(o={\rm nondet.} | \mathbf{s}) = \exp(-h(\mathbf{s}))(1- p_s).
\end{equation} 
That is, the number of detections in a timestep is treated as a Poisson process with rate $h$, and any number of detections $\ge1$ is considered equivalent. The factor $1-p_s$ enforces the fact that the agent observes the source if it finds it.  Note that the observations are defined so that $\sum_{o \in O} {\rm Pr}(o|s,a) =1.$ The detection likelihood for our choice of model parameters, with $\bar{S}=2.5,$ is shown in Fig.~\ref{fig:ellipses}.

Now, the problem becomes that of finding a good policy $\pi: b \mapsto a$
mapping each belief to an action which yields a maximal expected total reward (i.e., a short mean arrival time), conditioned on that belief. Explicitly, under a given policy $\pi$, we may define the \emph{value} $V_\pi$ of a belief as the total expected reward that can be accrued by following $\pi$:
\be V_\pi(b) = \mathbb{E} \left[ \sum_{t=0}^\infty \gamma^t \sum_{s\in S} R(s,\pi(b_t)) b_t(s) \bigg| b_0 = b\right]. \ee

We will define $V^*$ as the value function under the \emph{optimal} policy $\pi^*$. $V^*$ can be shown to satisfy the \emph{Bellman equation}
\be
\label{eq:bellman}
V^*(b) = \max_{a\in A} \left[ \sum_{s\in S} b(s) R(s,a) + \gamma \sum_{o\in O} {\rm Pr}(o|b,a) V^*(b_{o,a}) \right],
\ee
where ${\rm Pr}(o|b,a) = \sum_{s \in S} {\rm Pr}(o|s,a) b(s).$ Once a solution to the Bellman equations is found, the optimal policy consists in a greedy selection of the action that maximizes the RHS of (\ref{eq:bellman}). The argument of the maximum of the RHS of Eq.~\ref{eq:bellman} is simply the sum of the immediate expected reward for taking the action $a$ and the discounted expected reward for all future actions. Many solution methods for POMDPs are based on  ``value iteration'' on the Bellman equation, which is to say one computes
\be V^{n+1}(b) = \max_{a\in A} \left[ \sum_{s\in S} b(s) R(s,a) + \gamma \sum_{o\in O} {\rm Pr}(o|b,a) V^{n}(b_{o,a}) \right] 
\label{eq:perseusValueIteration}
\ee
until $V^n$ converges. However, due to the large size of the belief simplex, it is challenging to obtain an approximation which is good on a sufficiently large subspace of the belief simplex, and convergence may be slow. This is the ``curse of dimensionality'' and the fundamental issue making POMDPs hard.

\subsection{Initial belief}\label{sec:init}
While Bayesian inference suffices to specify the evolution of the agent's belief, we still need to set the initial belief $b_0$ that the agent holds when it starts searching (the prior). A na\"ive choice, common for many POMDPs, would be to start from a uniform belief on the grid. However, we argue this is unphysical, as insects in nature generally do not start searching unless they have detected a cue. Moreover, when we have tested a uniform prior, we find that the resulting policies have the agent tending to explore the full extent of the box in order to locate the boundaries.

A second idea, then, is to bias the uniform prior by enforcing artificially a detection at time $t=0$. This approach was taken in, for example, Ref.~\cite{loisy2021}. However, for our choice of parameters, detections are relatively rare except very close to the source. Thus, under this prescription, the agent will have the strong initial impression that it is within a few gridpoints away from the source, which is far from the ground truth.

\begin{figure}
    \centering
    \includegraphics[width=\linewidth,trim={0 1.5cm 0 1.5cm},clip]{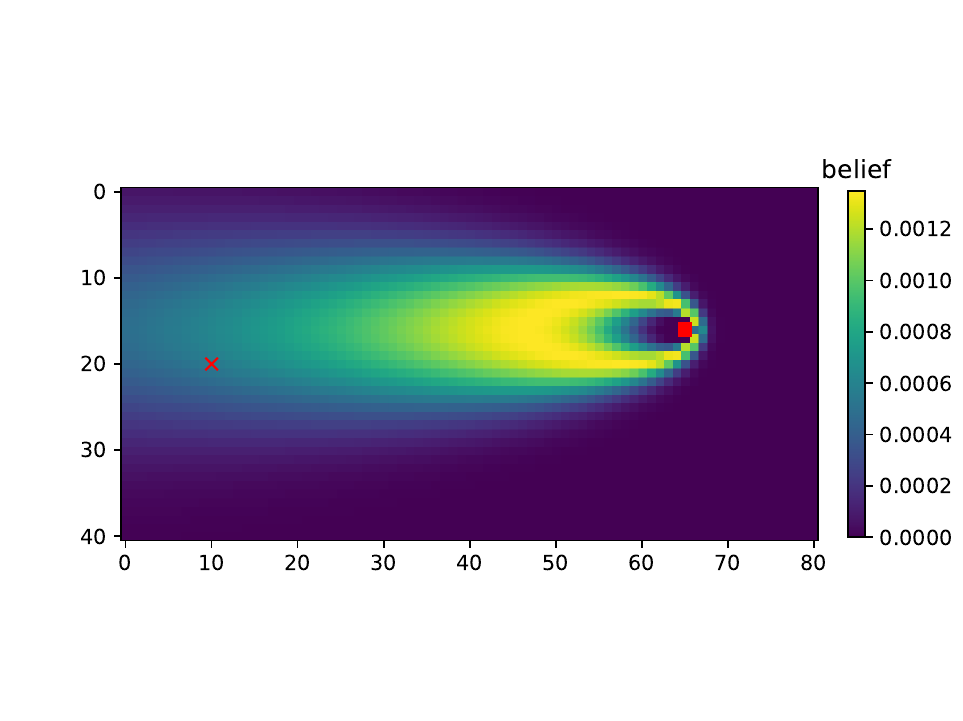}
    \caption{Plot showing a typical initial belief in an environment with emission rate $\bar{S}=2.5$. The agent location is indicated with a red square and the source location is indicated with a red X, and positions are measured in units of the grid spacing.}
    \label{fig:init_belief}
\end{figure}

Instead, we have elected to use a third approach, wherein the agent starts with a uniform prior and then waits in place for up to $T_{\rm wait}$ timesteps, continually updating its belief, until it makes a detection. Only after this detection does the agent begin searching. Thus the agent's initial belief is itself a random variable which carries some amount of useful information about where the source might be. This is intended to model the reasonable hypothesis that the insect knows the source is unlikely to be very close when it receives its first detection signal. A typical initial belief is shown in Fig.~\ref{fig:init_belief}.

To be clear, our choice of the initial belief is nothing more than a physical model, and it is not obvious that it should be preferred to the second approach (with a detection at time $t=0$). We will briefly explore this alternate initialization in Appendix \ref{app:init}.

\subsection{POMDP solution}
POMDPs are in principle exactly solvable by dynamic programming \cite{cassandra1998}, but due to the curse of dimensionality such an approach is usually so computationally expensive as to be intractable. Instead, approximation methods are generally preferred.

In this work, we use the Perseus algorithm \cite{perseus} to find near-optimal policies which approximately solve the Bellman equation. Perseus is a \emph{point-based value iteration} algorithm \cite{pineau2006,shani_review} for POMDPs, which means it involves collecting an initial large sample of beliefs ${\cal B}$ and then performing value iteration on those beliefs in order to obtain an approximation to the optimal policy. At each stage $n$ in the value iteration (cf.\ Eq.~\ref{eq:perseusValueIteration}),  $V^n$ is approximated by a piecewise-linear and convex function, represented by a collection $\mathcal{A}^n$ of hyperplanes in the belief simplex called $\alpha$-vectors. We have
\be
\label{eq:pwlc}
V^{n}(b)=\max_{\alpha \in \mathcal{A}^n} \alpha \cdot b,
\ee
where the dot product is over the states $s \in S$. Each $\alpha$-vector has an action associated with it, such that the (near-)optimal action for each belief is that associated with the maximizing $\alpha.$

The assembly of the belief set ${\cal B}$ can in principle be performed using any policy. We use infotaxis (see Sec.~\ref{sec:heuristic}) in this work.

We also accelerate the convergence of the algorithm using a \emph{reward shaping function} \cite{ng1999}. One can show (see Appendix for details) that adding a shaping function 
\be
F(s,a) = \phi(s) - \gamma \sum_{s'} p(s'|s,a) \phi(s') 
\ee
to the reward will not change the optimal policy, for any state-dependent function $\phi(s)$. We will take 
\be \phi(s) = -g(D(s)),\ee
where $g$ is some monotonically increasing function (with $g(0)=0$) and $D$ is the distance to the source, according to the metric induced by the state and action spaces --- here, the Manhattan distance. The point of this choice is to incentivize the agent to move closer to the source. We tested several such $g$ in this work, and found that, on problems of this size, a good choice of the shaping improves both the speed of convergence of Perseus and the performance of the resulting policies (see the Appendix for more details). In particular, we are unable to achieve comparable performance in the absence of reward shaping function. In contrast, on smaller problems with $O(100)$ points, we found that reward shaping was unnecessary and even counterproductive.

\subsection{Heuristic strategies}
\label{sec:heuristic}
As an alternative to trying to find a (near-)Bellman-optimal policy, one can instead propose a heuristic policy for a POMDP. Whereas solving the POMDP directly is a ``black-box'' approach that tries to directly maximize the reward over the entire horizon via (approximate) dynamic programming, a heuristic policy prescribes a simple, interpretable rule to choose an action, often by considering only a single time step in the future. Somewhat remarkably, there are a number of heuristics which can be effective for the search problem, despite its difficulty; here we present a few (see \cite{fernandez2006} for a review).
\subsubsection{QMDP}
Every POMDP has an underlying (fully-observable) MDP, for which the optimal policy $\pi^*$ is generally much easier to specify. One can then calculate the value of taking action $a$ in state $s$, the so-called Q-function 
\be
Q(s,a) = \mathbb{E} \left[ r(s,a) + \sum_{t=1}^\infty \gamma^t r(s_t,\pi^*(s_t))\right].
\ee 
For our search problem, the MDP-optimal policy is just to take the path of minimal distance to the source, so we have $Q_{\rm MDP} (s,a) = \gamma^{D(s')}$, where $s'$ is the state resulting from taking action $a$ in state $s$ and $D(s)$ is again the gridwise distance to the source (Manhattan distance). This observation motivates the \emph{QMDP} policy, which selects the action which maximizes the expectation of $Q_{\rm MDP}$ (conditioned on the belief):
\be
\pi_{\rm QMDP} (b) = \argmax_{a \in A} \sum_{s\in S} Q_{\rm MDP}(s,a) b(s) .
\ee

Because it tries to directly minimize the expected time to reach the source, the QMDP policy tends to be exploitative. We will see that it is only effective for this problem when the emission rate is relatively high. 

\subsubsection{Infotaxis and space-aware infotaxis}
The fundamental challenge of a POMDP is to make good action choices when faced with uncertainty. If the belief were perfectly informative (a $\delta$ distribution), then we would have a fully-observable MDP and the problem would be relatively trivial. This motivates an approach that tries to directly maximize the information content in the belief, or equivalently to minimize the Shannon entropy. Let 
\[ H[b] = -\sum_{s \in S} b(s) \log b(s), \]
with the logarithm expressed in some units of choice. Then we can craft a policy which chooses the action maximizing the immediate expected information gain,
\be
\pi_{\rm info}(b) = \argmin_{a \in A} \sum_{o\in O} {\rm Pr}(o|b,a) H[b_{o,a}].
\ee
In the context of olfactory search, this policy is called \emph{infotaxis} in analogy to chemotaxis \cite{infotaxis}. 

Since one of the possible observations is to find the source, which would collapse the belief into a $\delta$ distribution, the infotactic policy naturally balances the immediate reward of finding the source with longer-term rewards associated with exploration. However, the probability of finding the source immediately is usually small, so the explorative component tends to dominate; indeed, we will see infotaxis has an often excessive tendency towards safety. Thus, a number of variations and improvements have been proposed. One recent and promising variant, dubbed ``space-aware infotaxis'' (SAI), essentially combines infotaxis with the QMDP policy \cite{loisy2021}. Explicitly, the policy is
\be
\pi_{\rm SAI}(b) = \argmin_{a \in A} \sum_{o\in O} {\rm Pr}(o|b,a) \log_2 \left( \sum_{s \in S} D(s) b_{o,a}(s) + 2^{H_2[b_{o,a}]-1} + \frac12 \right).
\ee
We have chosen the base-2 logarithm and measured the entropy in bits to be consistent with the original authors, but our implementation differs very slightly in that we have reversed the sign in front of $1/2$, which ensures that the contribution to the outer sum from finding the source is nonsingular.

The second term of the summand is a crude estimation of the expected time to learn the location of the source (i.e.\ by checking $2^{H_2}$ cells), so the SAI policy is an attempt to directly minimize the total time to find the source. More generally, it balances infotaxis' tendency towards exploration with QMDP's tendency towards exploitation, and we will see it performs quite well.

\subsubsection{Thompson sampling}
A classical heuristic for decision problems with partial information is at, each timestep, to estimate the true state by \emph{sampling} from the current belief (posterior) and then to choose the action with maximal expected reward according to that sample. This strategy is usually referred to as \emph{Thompson sampling} \cite{thompson1933,thompson1935,russo2018}.

We adapt Thompson sampling to the search problem in the following way. At each timestep, we sample a possible source location from the current belief, and then take the action which brings the agent closest to that location (if there is more than one such action, we choose from these at random). 

We also find it useful to generalize this policy by introducing a persistence time $\tau \ge 1$: rather than sample a new location at every timestep, the agent follows the sampled location for $\tau$ timesteps (or until it reaches the sampled location), and only then does it resample. The benefit of pursuing a sample for an extended time rather than resampling at every timestep is linked to the need for ``deep exploration'' when navigating problems with sparse reward structure \cite{russo2018,osband2019}. 

That Thompson sampling depends on moving towards random locations suggests it should have a tendency towards exploration. Indeed, we will see it is generally a safe policy that, for sufficiently large $\tau,$ performs especially well in environments with low emission rate and thus low information. 

\section{Results}\label{sec:results}
\FloatBarrier
Once the Perseus algorithm has reached convergence, we freeze the policy and test it against the heuristic baselines described in Sec.~\ref{sec:heuristic}. We present results for three different model environments, which have emission rates $\bar{S}=0.25,2.5,$and $25$ but are otherwise identical. We suggest the most direct ways to change the character of the problem are to either alter the emission rate or move towards the windless, isotropic limit $V\lambda/D \to 0$, which we will not study here as it was examined thoroughly in Ref.~\cite{loisy2021} and will be further studied in Ref.~\cite{loisy2023}. The windy case is also arguably more relevant to insect behavior. The agent's model for the environment, by which it performs Bayesian updates, is exact, and detection events are drawn at random from the distribution defined by Eqs.~\ref{eq:det}--\ref{eq:nondet}. Results are shown for several choices of the shaping function $g(D(s)).$

The discount factor is a very important hyperparameter and we find values in the range $0.95\ge\gamma\ge0.99$ work best. In the environment with $\bar{S}=0.25$, we set $\gamma=0.96,$ and in the other two cases we set $\gamma=0.98$. Further optimization may be possible but is not the goal of this study. For additional details on $\gamma$ dependence, as well as other convergence properties of Perseus, refer to the Appendix.

\begin{figure}
    \centering
    \includegraphics[width=\linewidth]{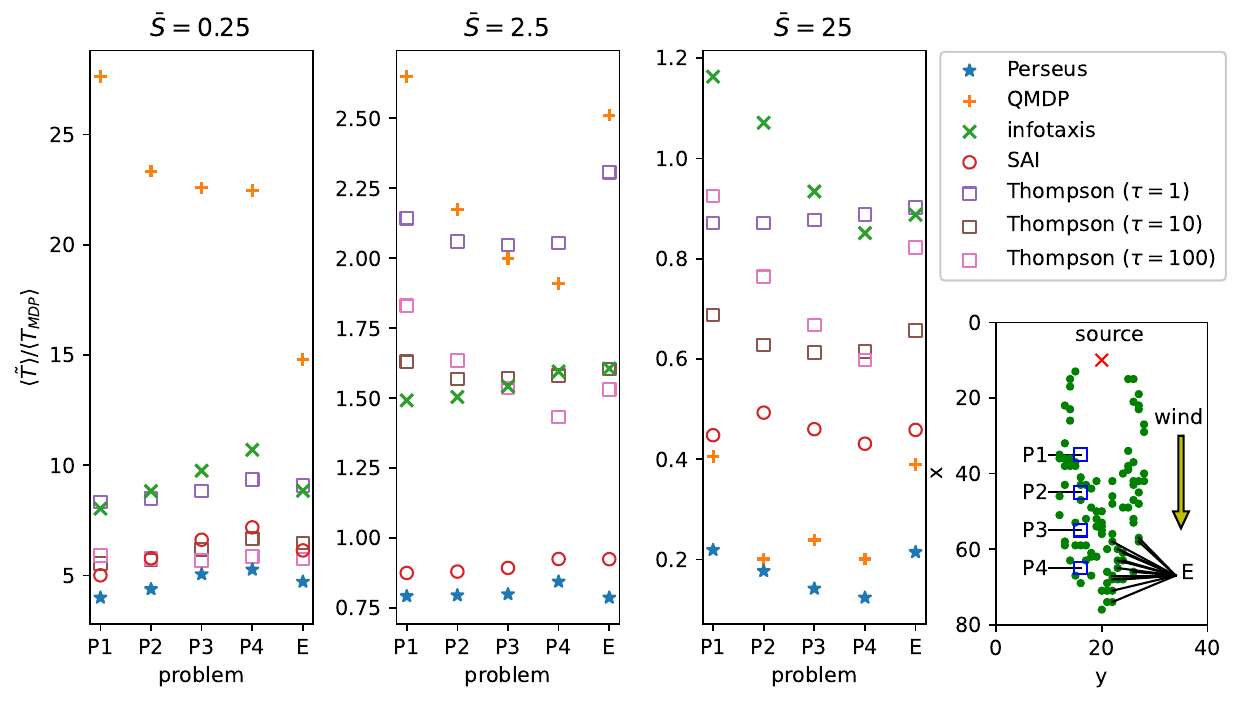}
    \caption{ Comparison of performance of Perseus (using the same policies as those chosen in Sec.~\ref{sec:pdfs}) vis-\`a-vis heuristic policies on the three environments tested. Problems P1-P4 represent starting from the single points $(25,-4),(35,-4),(45,-4)$ and $(55,-4)$ in order of increasing distance from the source, and problem E is the ensemble of starting points described in Sec.~\ref{sec:testing}. The performance on each problem is measured by the mean excess arrival time $\langle \tilde T\rangle \equiv \langle T \rangle - \langle T_{\rm MDP} \rangle,$ normalized by the minimum time $\langle T_{\rm MDP} \rangle.$ Error bars are suppressed for visual clarity; uncertainties on the means were at most 3.3\%, as measured by the standard error. In the lower right corner, we show the test problems: the ensemble E (for $\bar{S}=2.5$) as green circles, and P1-P4 as blue squares.}
    \label{fig:overall}
\end{figure}

The main results are summarized in Fig.~\ref{fig:overall}, where the performance of all policies are tested on both individual starting points and an ensemble of starting points. The individual points are Problems P1-P4 represent starting from the single points $(25,-4),(35,-4),(45,-4)$ and $(45,-4)$, where the coordinates are relative to the source and given in units of the grid-spacing. Performance is measured by the excess arrival time $\langle \tilde T\rangle \equiv \langle T \rangle - \langle T_{\rm MDP} \rangle,$ which we then normalize by the minimum time $\langle T_{\rm MDP} \rangle.$ For the single-point problems, we exclude ``failed'' trials with $T\ge 10000$ from the mean (in no case, except for QMDP when $\bar{S}=0.25$ or $\bar{S}=25$, was the failure rate significant), whereas, for the ensembles, we include the case $T\ge 1000$ (which all are registered as 1000 due to the time limited imposed). Anecdotally, failures are usually due to the agent somehow becoming trapped, and seem to be correlated with extremely rare events, i.e.\ when the agent makes a detection in a region where the likelihood is very low.

We see that Perseus clearly outperforms all tested heuristics, and that the best-performing heuristic depends strongly on the emission rate, making it difficult to select \emph{a priori} an optimal baseline. We can also conclude that space-aware infotaxis is a strictly better policy than its ``vanilla'' counterpart, that QMDP is inferior except when the emission rate is large, and that Thompson sampling performs best with a persistence time $\tau>1$ and when the emission rate is small. We briefly remark that in certain cases, especially when $\bar{S}=25,$ the normalized mean arrival time decreases with starting distance from the source. This is not a contradiction, and simply means that the excess arrival time is increasing sublinearly with distance.

\subsection{Excess arrival time pdfs from a single starting point}\label{sec:pdfs}
\FloatBarrier
Perseus was run on each environment using several choices of the reward shaping function. We then chose, for each emission rate, one Perseus policy and performed 20000 Monte Carlo trials from a single starting point, P3 --- $(45,-4)$. We did the same for the heuristic policies. Here, the agent was allowed up to 10000 timesteps to reach the source, to better resolve the tails of the distributions.

We selected for testing Perseus policies that (a) outperformed heuristics on the ensemble averaging (b) outperformed heuristics on the four individual points and (c) were evolved for as many iterations as possible.

\begin{figure}
    \centering
    \includegraphics[width=0.9\linewidth]{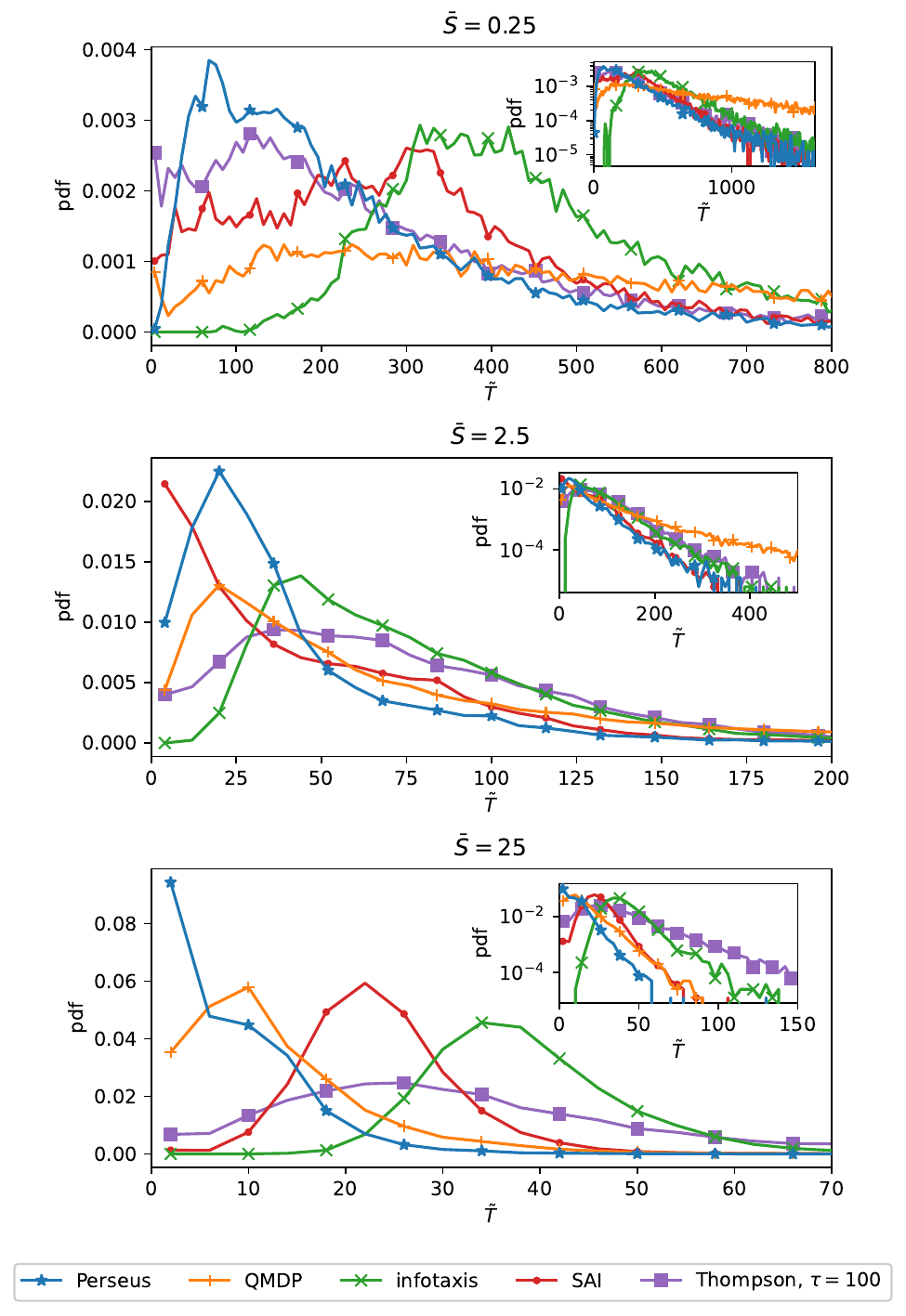}
    \caption{Comparison of excess arrival time pdfs, in each environment, from the point $(45,-4)$ for a selected Perseus policy and several heuristics. The insets show the same pdfs on a log-linear scale to emphasize the tails.}
    \label{fig:pdfs}
\end{figure}

The resulting pdfs are shown in Fig.~\ref{fig:pdfs}. Note that their ``wiggly'' appearance is due to finite sample size effects. These pdfs help illustrate the strengths and weaknesses of the heuristic policies. We see that infotaxis clearly tends to be too safe, with a tail that decays quickly for large arrival time but with a relatively small probability of rapidly arriving to the source. Infotaxis performs best at intermediate emission rate, but is clearly outperformed in all cases by SAI. SAI, too, is especially good at intermediate emission rate, and is closely competitive with Perseus. In fact, for intermediate emission rate, SAI's arrival time pdf peaks at the minimum arrival time; the Perseus policy only achieves a lower mean arrival time due to having substantially less probability of a long arrival time ($\tilde T\gtrsim 60$). QMDP is again seen to be inferior except when the emission rate is large. The excessive greediness of this policy can be seen by the very heavy, slowly-decaying tails when $\bar S = 0.25$ or $2.5.$ Finally, Thompson sampling shines at low emission rate, achieving results competitive with Perseus, but is otherwise relatively mediocre. 

\subsection{Sample trajectories and policy similarity for ${\bar S}=2.5$}
\begin{figure}
    \centering
    \includegraphics[width=\linewidth,trim={0 30pt 0 30pt},clip]{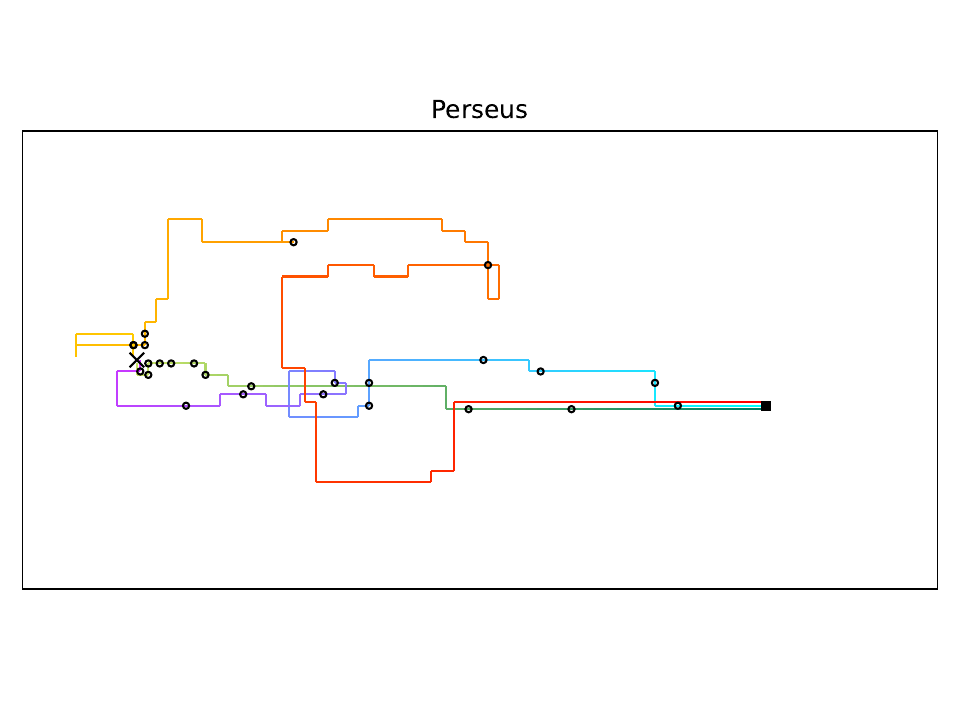}
    \includegraphics[width=\linewidth,trim={0 30pt 0 30pt},clip]{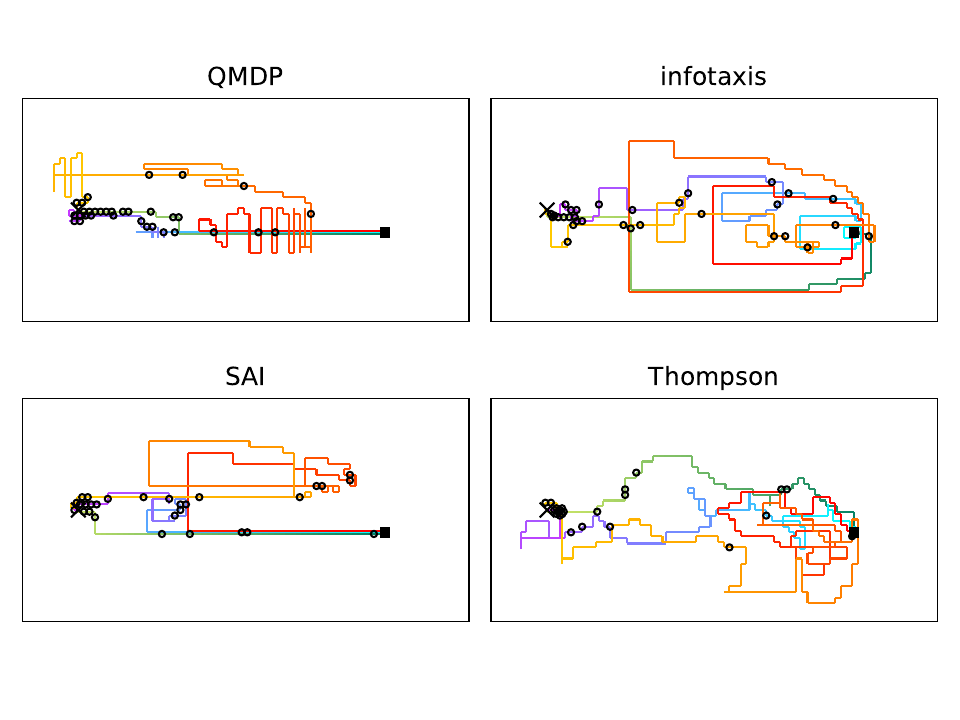}
    \caption{Sample trajectories from the same starting point using various policies in the $\bar{S}=2.5$ environment. The source is indicated with a black X, and the starting point with a red square. Detection events are indicated with black circles. We have deliberately chosen a trajectory with an especially small arrival time (green), a trajectory close to the median (light blue-purple), and a trajectory with an especially long arrival time (red-orange). The color gradients indicate the passage of time so that there is no ambiguity when the trajectories self-intersect. The trajectories have been offset in space slightly for visual clarity. We show trajectories for the near-optimal Perseus policies, QMDP (with $\gamma=0.98$), infotaxis, SAI, and Thompson sampling (with $\tau=10$).}
    \label{fig:traj}

\end{figure}

\begin{figure}
    \centering
    \includegraphics[width=\linewidth]{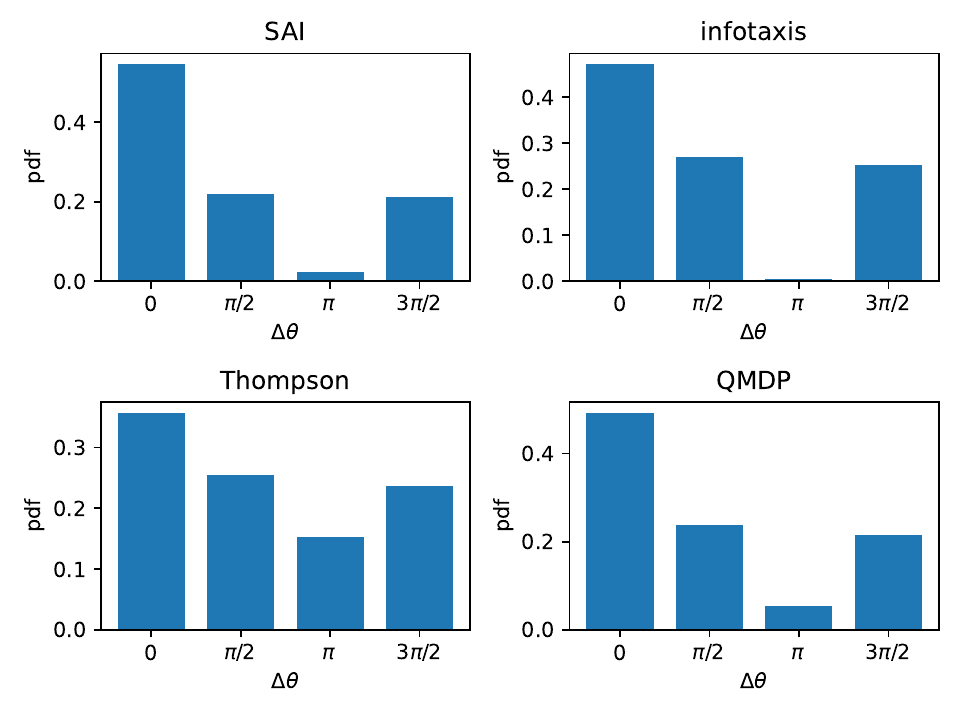}
    \caption{Pdfs for angular differences between Perseus policy and other policies, when $\bar{S}=2.5.$ Here, the Thompson sampling policy used $\tau=10$ and the QMDP policy used $\gamma=0.98.$}
    \label{fig:angles}
\end{figure}

\FloatBarrier
In Fig.~\ref{fig:traj}, we show some sample trajectories in the intermediate-emission-rate environment ($\bar{S}=2.5$). In Fig.~\ref{fig:angles}, we compare policies to Perseus in this environment by estimating the pdfs for angular differences in chosen actions. To be precise, we let the agent search using the Perseus policy, and at each timestep we also compute the actions which would have be chosen by the heuristic policies, given the same belief. These actions are converted into a polar angle $\theta\in \{0,\pi/2,\pi,3\pi/2\},$ and we record the angular differences (modulo $2 \pi$) $\Delta \theta= \theta_i - \theta_{\rm Perseus}$ between the Perseus action and the heuristic actions for each policy $i$. The starting points were selected randomly in the usual way, and a thousand Monte Carlo trials were performed. Unsurprisingly, of the four heuristics tested, SAI was most similar to the near-optimal policy, with over 50\% of the actions being identical.

The trajectory plots are of course qualitative, and a careful quantitative analysis of the behaviors which typify different policies is beyond the scope of the present work. However, it is worth commenting briefly on some of the observed features. In general, each policy tends to result in a diverse set of behaviors, and policies are not always easily distinguishable by the naked eye. Nevertheless, a few traits are apparent. The Perseus agents tend to prefer to move upwind initially before beginning crosswind motion. If some time passes without a detection, there is often a tendency to return in the downwind direction, a behavior which helps prevent the agent from overshooting the source. Infotaxis agents often move in broadly arcing trajectories, including outward-moving spirals. Spiraling behavior under infotaxis is well-known, having been observed in the original paper \cite{infotaxis} as well as subsequent work \cite{masson2009,barbieri2011,loisy2021}, and it can be connected to the theory of search games: one can show that the optimal minimax trajectory for searching for a single point in a plane is an exponential spiral \cite{alpern2003}. SAI trajectories are often similar to the Perseus trajectories, sharing the tendency to initially move upwind (a consequence of the distance term in the objective function). QMDP tends to be a greedier policy; QMDP agents have a tendency to either strike downwind or search semi-exhaustively on small scales. Thompson sampling agents exhibit meandering trajectories characteristic of the randomness of the underlying policy; this randomness is also consistent with the relatively small observed probability that a Thompson agent will chose the same action as a near-optimal Perseus agent. As a final remark, we found that all policies tested in the present work sometimes yielded behaviors which resemble casting. 

\subsection{Dependence of problem difficulty on starting point}
\FloatBarrier
How does the mean arrival time of a (near-)optimal policy depend on the starting position? Trivially, the arrival time is bounded below by the MDP optimum, which measures the component of the problem difficulty due to the time needed to reach the source. The mean excess arrival time $\langle \tilde T \rangle $ then measures, in a sense, the component of the difficulty due to partial observability, or the time required to gather information and determine where the source is. It is not immediately obvious how $\langle \tilde T \rangle $ should depend on the starting position.

\begin{figure}
    \centering
    \includegraphics[width=\linewidth]{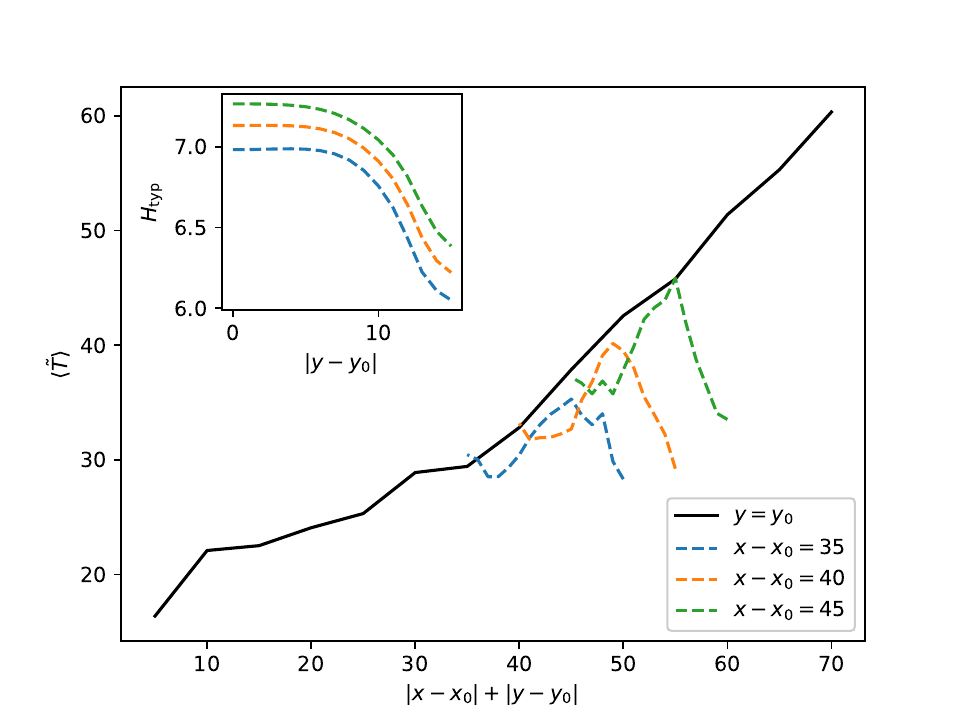}
    \caption{Dependence of the mean excess arrival time on the initial Manhattan distance from the source, using the near-optimal Perseus policy, for $\bar{S}=2.5$. We compare curves at fixed downwind distances $x-x_0$ to the curve at fixed $y=y_0$ (i.e., where the agent starts on the symmetry axis). The inset shows typical entropy values $H_{\rm typ}$ for the fixed downwind distance curves.}
    \label{fig:downwind}
\end{figure}

To help answer this question, in Fig.\ref{fig:downwind}, we plot the mean excess arrival time using Perseus, when $\bar{S}=2.5$, as a function of the downwind distance at fixed crosswind distance, and vice-versa. The averages were taken over $10^4$ Monte Carlo trials, and arrival times greater than or equal to 5000 were suppressed from the calculation. 
Somewhat surprisingly, the problem difficulty does not appear to depend strongly on the starting crosswind distance, as long as the agent starts within a few crosswind diffusion lengths from the symmetry axis. Instead, the excess arrival time mostly depends on the downwind distance, scaling approximately linearly therewith. From Fig.~\ref{fig:downwind} it is clear that starting further downwind from the source generally makes the problem more difficult that starting further off-axis.

This behavior can be qualitatively explained by the following argument. The problem of finding the source can be solved by locating the symmetry axis (say, by casting) and then proceeding upwind. Assuming that $\tau$ is large, a detection at $(x,y)$ (measured with respect to the source) means that $y^2 < 2 \lambda x$ with high probability, since outside of this parabola, detections are exponentially suppressed with decay lengthscale $\lambda$. The further away from the source the agent starts, the wider this parabola is, and the wider the agent must cast in order to make a detection (which of course consumes more time). On the other hand, the time spent locating the symmetry axis in a casting-based strategy should \emph{not} depend on its initial crosswind distance from the axis. 

We also note a precipitous drop in the excess mean arrival times with crosswind distance starting around $y=10$. This can be explained by a drop in the entropy of a typical starting belief when one starts sufficiently far from the symmetry axis. If the agent starts at $\mathbf{r}_i,$ the time to the first hit, call it $k$, obeys a geometric distribution with parameter $\ell_0=\mathrm{Pr}(o={\rm hit}|\mathbf{r}_i-\mathbf{r}_0)$, and we have $\mathbb{E}[ k ] = 1/\ell_0$. If $\ell(s)$ is the probability of detection as a function of states, the belief will then accrue, through Bayesian updates, $k-1$ factors of $1-\ell(s)$ (one for each non-detection) and one factor of $\ell(s).$ Thus a typical initial belief is
\be
\label{eq:typical_belief}
b(s)= \frac{(1-\ell(s))^{1/\ell_0-1} \ell(s)}{\sum_s (1-\ell(s))^{1/\ell_0-1} \ell(s)}.
\ee
The entropies of the typical beliefs corresponding to the starting positions of the constant downwind distance curves of Fig.~\ref{fig:downwind} are shown in the inset of that figure. The reason for the entropy drop is restriction of the support of the initial belief to a progressively smaller area: as $k$ gets larger, the initial belief is confined to the maximum likelihood curve $\ell(s) = 1/k$ with increasingly narrow characteristic width $w\propto 1/\sqrt{k-1}$.

\subsection{Robustness of policies to changes in environment}
Because POMDP is a model-based approach, we have assumed that the insect has some instinctual knowledge of the turbulent environment. Up until now, this knowledge has been an exact model of the detection statistics. In this section, we relax this strong assumption and experiment with a scenario where the agent has an \emph{imperfect} model of the environment. In particular, the true physical parameters will be different from those the agent uses to update its belief and those which were used to construct a near-optimal policy. We present results for two cases: one, a more turbulent environment where $D \to 2 D$ and $V \to V/2$, and two, a less turbulent environment where $D \to D/2$ and $V \to 2V$, relative to the parameters we have used previously. The agent will use the old parameters to update its belief. In both cases, we set $\bar S = 2.5$. 

\begin{figure}
    \centering
    \includegraphics[width=\linewidth]{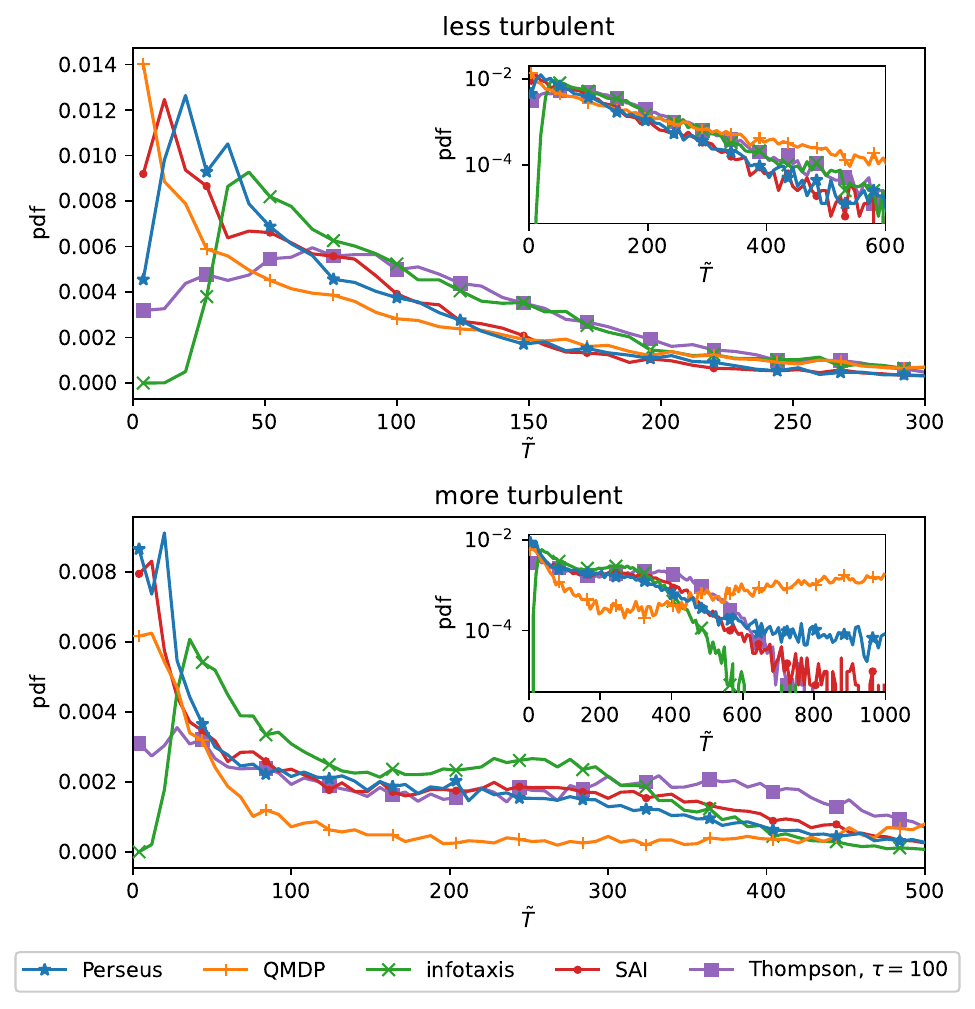}
    \caption{Arrival time pdfs from $(45,-4)$ for the previously obtained near-optimal policy for $\bar{S}=2.5$ and for heuristics, but now searching in environments which are less (top) and more (bottom) turbulent than the agent believes.}
    \label{fig:more_turb}
\end{figure}

In Fig.~\ref{fig:more_turb} we show excess arrival time pdfs obtained using the same methods as in Sec.~\ref{sec:pdfs}, but now for these two scenarios. The mean excess arrival times are shown in Table~\ref{table:robustness}, with the previous results where the model is exact shown for comparison. In Table~\ref{table:failure}, we show failure rates for the problems.

In the more turbulent environment, the Perseus policy performs poorly compared to most other heuristics, in particular suffering a high rate of failure and a fat tail, leading to a large mean arrival time. Infotaxis has the best performance here, due to its rapidly decaying tail, which is consistent with its being a ``safe'' policy. Moreover, whereas the failure rates were negligibly small previously, Perseus and QMDP now have substantial probabilities of failing (i.e.\ taking $\ge 10^4$ timesteps to reach the source). 

The problem of searching in an environment that is less turbulent than believed is evidently substantially easier. In the less turbulent environment, the Perseus policy performs very adequately, scoring the second best performance behind SAI. Perseus also has a substantially smaller failure rate on this problem than in the more turbulent environment. This underscores an obvious drawback of training a policy to be optimal for a given environment: if the environment changes, the policy may be too overtuned to perform adequately. However, these results suggest that if the environmental parameters are unknown, for the sake of robustness it may preferable to train a policy to be optimal assuming a more turbulent environment.

\begin{table}[]
    \centering
    \begin{tabular}{|c|c|c|c|}
    \hline
       policy & $\langle \tilde T \rangle$ (E) & $\langle \tilde T \rangle$ (MT) & $\langle \tilde T \rangle $ (LT) \\
       \hline
       Perseus & $39.1 \pm 0.3 $ & $349.1\pm 6.2$ & $91.2 \pm 1.6$ \\
       QMDP & $97.9 \pm 1.4 $ & $1852.1\pm 11.1$ & $231.4 \pm 4.4$ \\
       infotaxis & $75.5 \pm 0.3$  & $174.5\pm 0.9$ & $120.1 \pm 5.9 $\\
       SAI & $ 43.8 \pm 0.3$ & $179.4 \pm 1.2$ & $79.6 \pm 0.6$ \\
       Thompson ($\tau=10$) & $77.0 \pm 0.3 $ & $262.1\pm1.3$ & $105.2 \pm 0.5$ \\
       \hline
    \end{tabular}
    \caption{Mean excess arrival times from $(45,-4)$ (with standard error shown) for $\bar{S} = 2.5$ on three problems: the original problem where the agent's model for the environment is \emph{exact} (E), a scenario where the true detection statistics are reflective of a \emph{more turbulent} environment (MT) than the agent believes, and  a scenario where the true detection statistics are reflective of a \emph{less turbulent} environment (LT) than the agent believes.}
    \label{table:robustness}
\end{table}

\begin{table}[]
    \centering
    \begin{tabular}{|c|c|c|c|}
    \hline
       policy & failure rate (E) & failure rate (MT) & failure rate (LT) \\
       \hline
       Perseus & $5\times 10^-5$ & 0.0264 & 0.0017  \\
       QMDP & $5\times 10^-5$ & 0.00935 & 0.0096 \\
       infotaxis & 0 & $10^{-4}$ & 0\\
       SAI & 0 & 0 & 0 \\
       Thompson ($\tau=10$) & 0 & 0 & 0 \\
       \hline
    \end{tabular}
    \caption{Same as Table \ref{table:robustness}, but showing failure rates for the problems.}
    \label{table:failure}
\end{table}

\section{Discussion}\label{sec:discussion}
We have studied a search problem relevant to the behavior of a number of flying insects. We have shown that solving for near-optimal search policies on a problem space with several thousand points is feasible using an existing POMDP algorithm, Perseus, when accelerated with a good choice of reward shaping. This approach yielded policies which outperformed all tested heuristics in terms of the mean arrival time, over a wide range of emission rate regimes. We are thus optimistic that more sophisticated search problems, such as ones which take spatiotemporal correlations between detections into account, could be amenable to a direct POMDP solution approach, despite necessitating a larger POMDP state space. Future work will investigate such problems.

We also studied which heuristic strategies perform best in environments with different characteristic concentration levels (emission rates). In particular, we found a randomized search algorithm --- Thompson sampling --- to be well-suited for very dilute environments, space-aware infotaxis to be excellent at a somewhat higher concentration, and QMDP to be effective only on the easiest problems with substantial detection rates. It should be noted, however, that these conclusions may be sensitive to the choice of prior.

An advantage of certain heuristics, especially Thompson sampling and variants of infotaxis, over a near-optimal policy is their being more flexible when the agent's model of the environment is imperfect. Finding policies which are effective in a variety of environments is an interesting avenue of future research; in that case, model-free approaches may be preferable to POMDP. 

We found that a variety of behaviors emerge from different policies, but classifying these carefully is highly nontrivial since behaviors depend on the observation history and reflect correlations between actions over relatively long time scales. For instance, we tried measuring the fraction of time each policy spent moving crosswind, upwind, or downwind as a simple metric but did not find it informative. Thus, we defer a serious quantitative study of behaviors to future efforts.

It should be noted that while we found increases to the emission rate led to a reduction in typical arrival times, this trend cannot continue indefinitely. If the emission rate is sufficiently large, then the likelihood of detection will begin to saturate near unity and a binary detection scheme will cease to be informative or useful. Instead, in this regime one would expect gradient-based (chemotactic) strategies to once again be effective.

Finally, we used a near-optimal policy to study the spatial dependence of the mean excess arrival time, a proxy for the intrinsic problem difficulty as a function of starting point. This dependence is strongly anisotropic: the mean excess arrival time increases monotonically as the agent starts further downwind, but has a strongly non-monotonic dependence on crosswind distance. Moving only a few $\lambda$ off-axis has virtually no effect on the problem difficulty and may even make it slightly easier, which may be related to why cast-and-surge is an effective search strategy.

The approach proposed in this work can be extended in several directions. First, and most importantly, it would be extremely interesting to validate it on realistic data from direct numerical simulations of emission from a point source in 2--D or 3--D turbulent flows, where a Markovian model for observations will necessarily be incomplete. A study of this kind is currently ongoing. Second, similar techniques can be used to attack multi-source problems and/or multi-agent problems. It is increasingly urgent to identify clear set-up with high quality and quantity of data for training and validations of data-driven algorithms, and the one here studied is certainly a good paradigmatic candidate.

In this work, we kept the discount factor $\gamma$ as close to one as possible and aimed to minimize the mean arrival time $T$. Strictly speaking, however we were actually maximizing $\gamma^T;$ in general the closer $\gamma$ is to 1, the more we care about optimizing the tail of the arrival time pdf, which is to say avoiding very long arrival times. We have also noticed that in the presence of a nonzero failure rate, it is not always obvious which policy is ``best'' --- depending on ones tolerance of failure, a lower mean arrival time with higher failure rate may be preferable to a ``safer'' policy which almost always finds the source. These ideas can be generalized by the notion of \emph{risk-sensitive} problems \cite{howard1972}, where the objective function is transformed in a way that reflects the agent's aversion or attraction to risky behavior. Obtaining optimal policies for the search problem subject to risk-sensitivity is another subject for future study.

A forthcoming study \cite{loisy2023} will compare the performance of the deep-RL method proposed in \cite{loisy2021} to the approach of the present work on problems of this size. 

\appendix
\section{Detailed methods}\label{app:methods}
\subsection{POMDP implementation}
Defining the POMDP in a careful way is important to keep the problem tractable and to avoid issues with boundary artifacts. Hence, perhaps at the risk of seeming pedantic, we will try to be precise as possible in what follows.

Let $G$ be the $N_x \times N_y$ gridworld. The POMDP state space may be defined to be the Cartesian product of the possible agent locations $\mathbf{r}$ and the possible source locations $\mathbf{r}_0$, $G\times G$, and we have a belief $b$ on the $G \times G$-simplex which is the joint distribution of the agent location belief and the source location belief. This belief simplex is intractably large, $(|G|^2-1)$-dimensional. However, we can exploit to our advantage (a) the sparsity of the agent location belief, which is a $\delta$ distribution since the agent knows where it is, and (b) the fact that the observation likelihoods depend only on the displacement $\mathbf{s}=\mathbf{r}-\mathbf{r}_0$ between the agent and source, and not on $\mathbf{r}$ or $\mathbf{r}_0$ independently.  

We map the state (living on $G\times G$) to a smaller state space, the $(2N_x-1)\times(2N_y-1)$-dimensional grid spanning $-L_x \le x \le L_x$ and $-L_y \le y\le L_y$, with spacing $\Delta x$ and $\Delta y$, which we call $G'$. $G'$ is the space of all possible displacements, given that the agent and source are both on $G$. We thus form a new belief $b'$ of displacements living on the $G'$-simplex. $b'$ is found by embedding the belief of source locations (the nonzero slice of the full belief) using the rule
\be
\label{eq:embedding}\begin{cases}
b'(\mathbf{s}) = b(\mathbf{r}-\mathbf{s}),& \mathbf{r}-\mathbf{s} \in G \\
0, & {\rm otherwise}.
\end{cases} \ee for all $\mathbf{s} \in G',$ where, in a slight abuse of notation, $b$ is the belief of the source location. This embedding can be inverted to recover $b$ from a given $b'$ using the known agent location $\mathbf{r}$.

Let us summarize the procedure. As the agent moves, it maintains a belief of the location of the source location $b$. $b$ is updated after taking an action $\mathbf{a}$ and making an observation $o$ according to Bayes' rule
\begin{equation}
\label{eq:full_update}
b(\mathbf{r}_0|o,\mathbf{a}) =  \frac{b(\mathbf{r}_0) \mathrm{Pr}(o|\mathbf{r}',\mathbf{r_0})}{\sum_\mathbf{r_0} b(\mathbf{r}_0) \mathrm{Pr}(o|\mathbf{r}',\mathbf{r_0})},
\end{equation}
where $\mathbf{r}'$ is the position of the searcher after previously being at $\mathbf{r}$ and taking action $\mathbf{a}.$ Explicitly,
\be
\begin{cases}
\mathbf{r}'=\mathbf{r}, & \mathbf{r}+\mathbf{a} \notin G \textrm{ or } \mathbf{r}=\mathbf{r}_0 \\
\mathbf{r}'= \mathbf{r}+\mathbf{a}, & \textrm{otherwise}.
\end{cases}
\ee
Finally, the belief is embedded into the $G'$-simplex to form $b'$. The computed, near-optimal policies in this work use $b'$ as input. For the purposes of computing policies, the boundaries of $G'$ (which are generally not encountered by the agent) are chosen to be doubly-periodic for simplicity. 

The primary reason for choosing this somewhat complicated representation is that it avoids boundary artifacts while maintaining a relatively compact dimensionality (many fewer dimensions than the full belief on $G\times G$). Na\"ive implementations which depend on propagating the belief of the relative position of the agent can introduce such artifacts when the agent moves, causing probability mass to exit the domain and be lost. In our representation, we propagate the belief of where the \emph{source} is; this propagation does not introduce artifacts because the source is static and therefore the transition matrix for the source location is trivial: ${\rm Pr}(\mathbf{r}_0'|\mathbf{r}_0,\mathbf{a}) = \delta_{\mathbf{r_0},\mathbf{r_0}'}$ for any $\mathbf{a}\in A.$ 

\subsection{Alternate reward structures}\label{app:rewards}

A more general way to model the search problem as a POMDP would give a (discounted) penalty at each time step until the agent finds the source, and then supply some onetime reward $R\ge 0$ (in the case $R=0$ then we are directly minimizing the arrival time). But as long as $\gamma<1$, this is equivalent to the reward we used in the present work. If we let the arrival time be $T$, the reward is then 
\begin{align} \mathbb{E} \left[\sum_{t=0}^{T-1} (-1) \gamma^t + R\gamma^T\right] &= \mathbb{E} \left[- \frac{1- \gamma^{T}}{1-\gamma}  + R\gamma^T \right] \nonumber \\ &= \frac{1}{1-\gamma} ((1+R)\mathbb{E}[ \gamma^T]-1).
\end{align}
Thus maximizing the expected reward for this reward structure is equivalent to maximizing the expected reward for the structure with no penalty per unit time. 

It is an important technical point that, while our ultimate goal is to reach the source in a minimal time, we are not directly minimizing the mean arrival time $\langle T \rangle$, but rather maximizing a proxy $\langle \gamma^T \rangle.$ Note that in the limit $\gamma \to 1$, the two objectives are equivalent (provided that the typical $T$ does not diverge in this limit), as can be shown by simple Taylor expansion of $\gamma^T$.

We are thus effectively treating $\gamma$ as a hyperparameter of the algorithm, and we will simply select the one that yields the best empirical performance, as measured by the mean arrival time, on a given problem setup. In contrast, the more typical viewpoint would be to consider $\gamma$ to be a parameter which reflects the environment and/or the agent and its priorities, each value of which defines a separate problem with a different optimal policy.

The basic reason for our choice to use $\gamma<1$ is that most available POMDP algorithms, Perseus included, require use of a discount factor in order to converge. A key underlying fact is that when $\gamma<1,$ the right-hand side of the Bellman equation acts as a contraction operator on the value function, which guarantees the convergence of iterative solution techniques to a unique fixed point \cite{denardo1967}. 

The undiscounted case $\gamma=1$ results in direct minimization of the mean arrival time $\mathbb{E}[T]$ and is thus an interesting alternative problem. However, there are few algorithms which can deal with the absence of the discount factor, the DRL approach of Ref.~\cite{loisy2021} being a rare example. We have directly verified \cite{loisy2023} that introducing $\gamma$ does not substantially increase the mean arrival time, relative to that of the solution to the undiscounted problem.

\subsection{Reward shaping}
In many (fully-observable) MDPs, it is often possible to speed up convergence to an optimal policy by adding a well-chosen \emph{shaping function} to the reward. In particular, there exist \emph{potential shaping functions} such that the MDP under the transformed reward has precisely the same optimal policy as the original MDP \cite{ng1999}. This idea can be generalized to POMDPs in a straightforward manner.

We start by introducing the function $Q(b,a)$ which expresses the value (expected total reward) of taking action $a$ when the agent has belief $b$. In particular, we have
\be V^*(b) = \max_{a \in A} Q(b,a) \ee
and 
\be \pi^*(b) = \argmax_{a} Q(b,a). \ee
Importantly, the optimal policy is unchanged under any transformation $Q \to Q + \varphi(b).$ Intuitively, this means the component of the value of a state which is intrinsic to the state, independent of the choice of action, should not affect the policy. $Q$ satisfies its own Bellman equation
\be Q(b,a) = \sum_{s\in S} b(s) R(s,a) + \gamma \sum_{o\in O} {\rm Pr}(o|b,a) \max_{a' \in A} Q(b_{o,a},a'). \ee
Letting $\hat Q(b,a) = Q(b,a) + \varphi(b)$ and substituting, we have
\be
\hat{Q}(b,a) = \sum_{s} b(s) R(s,a) + F(b,a) + \gamma \sum_o {\rm Pr}(o|b,a) \max_{a'} \hat{Q}(b_{o,a},a'),
\ee
with
\be
F(b,a) = -\varphi(b) + \gamma \sum_o {\rm Pr}(o|b,a) \varphi(b_{o,a}),
\ee
which is a new Bellman equation for $\hat Q$. 

As an important special case, we can restrict $\varphi$ to a linear functional $\varphi(b) = \sum_s b(s) \phi(s)$. (This special case is especially useful for us because we seek a piecewise linear approximation for $V.$) Then the introduction of the potential is equivalent to modifying the reward $R(s,a) \to R(s,a) + F(s,a),$ where 
\be
F(s,a) = \phi(s) - \gamma \sum_{s'} p(s'|s,a) \phi(s'). 
\ee
Thus, adding any function of this form to the reward in a POMDP will not change the optimal policy. This flexibility in defining the reward is akin to a kind of gauge invariance. Note that if $\hat{V}^*$ is the value function under the shaped reward, we also have the simple identity 
\be\label{eq:shapedvalue} \hat{V}^*(b) = V^*(b) + \varphi(b). \ee

How do we choose a good shaping function? One can argue that the ``best'' potential, which would accelerate value iteration as much as possible, would in fact be, up to an additive constant, the (negative) optimal value function, which is suggested by Eq.~\ref{eq:shapedvalue}. To see this, consider value iteration in the presence of a generic shaping $\varphi(b)$:
\be
V^{n+1}(b) = \max_{a \in A} \left[ \sum_{s \in S} b(s) R(s,a) - \varphi(b) + \gamma \sum_{o \in O} {\rm Pr}(o|b,a) \varphi(b_{o,a}) + \gamma \sum_o {\rm Pr}(o|b,a) V^n (b_{o,a}) \right].
\ee

Suppose $V^0(b) = 0$ for some $b$ (if $\gamma<1$, we can always define the reward in such a way that 0 is a safe choice for the initialization of the value function), and suppose we choose $\varphi(b) = -V^*(b)$. Then value iteration would yield
\be
V^{1}(b) = \max_{a \in A} \left[\sum_{s \in S} b(s) r(s,a) -V^*(b)+ \gamma \sum_{o \in O} {\rm Pr}(o|b,a) V^*(b_{o,a})  \right]= 0,
\ee
where we have used the Bellman equation. The value at $b$ has already converged, and in particular the maximizing action is none other than $\pi^*(b)$. Thus, a single value iteration would instantly give the correct action in a neighborhood of $b$, which would reduce the problem to simply sampling enough $b$. 

Of course, we do not have access to the optimal value function (this would defeat the purpose of value iteration!), so we must make an inspired guess for the reward shaping function which is structurally similar to the true value. For the search problem, a natural choice is to shape the reward to encourage moving toward the source by setting, as we have in this work,
\be \varphi(b) = - \sum_{s} b(s) g(D(s)),\ee
where $g$ is monotonically increasing and $g(0)=0$, and $D$ is the Manhattan distance to the source.

The importance of ensuring the shaping function is potential cannot be overstated. A non-potential choice like rewarding the agent for making detections, while having intuitive merit, would explicitly destroy the optimality of the solution with respect to the arrival time.

\subsection{Perseus algorithm}

Value iteration in Perseus is accomplished through the \emph{backup} operation. With the aid of Eq.~\ref{eq:pwlc}, we can express the value iteration as
\be
V^{n+1}(b) = \max_{a\in A} \left[b \cdot R_a + \gamma \sum_{o \in O} \max_{i} b \cdot g^i_{a,o} \right],
\ee
where $R_a=R(s,a)$ and
\be
g^i_{a,o} = \sum_{s'\in S} {\rm Pr}(o|s',a) {\rm Pr}(s'|s,a) \alpha_i(s').
\ee 
There is a $g^i_{a,o}$ for each $\alpha$-vector, which can be computed once and stored. The backup operation is then
\be
\label{eq:backup}
\mathrm{backup}(b) = \argmax_{\{\alpha'_a\}_{a\in A}} b \cdot \alpha'_a
\ee
where
\be
\alpha'_a = R_a + \gamma \sum_{o \in O} \argmax_{\{g^i_{o,a}\}_i} b \cdot g^i_{o,a}.
\ee

The backup thus produces a new alpha vector $\alpha'$ so that $V^{n+1}(b) = b \cdot \alpha'$, and the optimizing action is found during the argmax in Eq.~\ref{eq:backup}.

The backup operator forms the basis of an array of different ``point-based'' algorithms \cite{shani_review}, which differ in how they sample beliefs from the simplex and in what order they are backed up. In the original Perseus algorithm, backups are performed in a random order, but we perform them in the order of decreasing \emph{Bellman error} --- the so-called ``prioritized'' version of Perseus. The Bellman error of a belief $b$ is defined as 
\be
\label{eq:bellmanerror}
\epsilon(b) = \max_{a \in A} \left[ b \cdot R_a + \sum_{o \in O} {\rm Pr}(o|b,a) V^n(b_{o,a}) \right] - V^n(b).
\ee
Prioritizing Perseus in this way was shown to accelerate convergence in \cite{shani2006}.

Our implementation of the Perseus algorithm proceeds as follows. First, we assemble a large collection of beliefs ${\cal B}$, by exploring the environment according to some policy (after initializing the agent in the same way that we do during evaluation), updating the belief using Bayes' theorem, and adding the belief to ${\cal B}.$ When the agent finds the source, we restart and repeat, until we have enough beliefs. The original paper \cite{perseus} suggested using a uniform random policy, but we find it is much better to employ a heuristic, as suggested in \cite{shani_review}. Intuitively, the most useful beliefs to sample are those which are likely to be encountered when taking optimal actions \cite{sarsop}, and it is generally understood that the subspace of these ``reachable beliefs'' is much smaller than the whole simplex. Thus using a good heuristic which is reasonably close to optimal is a far more efficient way to sample beliefs; we use infotaxis. We found that $|{\cal B}| = O(10^4)$ beliefs were required to obtain good results, with clear loss in performance when fewer beliefs were sampled.

Next, we initialize the policy to a single $\alpha$-vector: ${\cal A}^{0} = \{ \mathbf{0}\}$, the zero-vector. This was chosen to guarantee $V^0 (b) \le V^* (b)$ $\forall b$ (note that reward shaping does not affect this choice, as long the potential $\varphi(b)$ is non-negative, in view of Eq.~\ref{eq:shapedvalue}). We assign one of the four actions to this vector; it does not matter which.

Finally, we perform some number of iterations until a convergence or performance criterion is met. An iteration consists of the following steps:
\begin{enumerate}
    \item Compute, for each $b \in {\cal B},$ $\epsilon(b)$.
    \item Initialize ${\cal B}'$ to ${\cal B}$ and ${\cal A}^{n+1}$ to $\emptyset$.
    \item Find $\alpha = \mathrm{backup}(b)$ for the $b\in {\cal B'}$ with largest $\epsilon$. 
    \item If $\alpha \cdot b \ge V^n(b),$ then add $\alpha$ to ${\cal A}^{n+1}.$ Otherwise, add the $\alpha \in {\cal A}^n$ which previously gave the maximum value for $b$. 
    \item Set ${\cal B}' \gets \{ b \in {\cal B'}: \alpha \cdot b < V^n(b) \}$, where $\alpha$ is the vector added to ${\cal A}^{n+1}$ in step 4.
    \item If ${\cal B}' \ne \emptyset$, go to step 3. 
\end{enumerate}

Perseus is not the only algorithm for POMDP planning. We note in particular the existence of ``heuristic search value iteration'' (HSVI) \cite{hsvi,hsvi2} and SARSOP \cite{sarsop}, both of which involve building a tree of beliefs reachable from some (single) initial belief, and both of which come with more rigorous guarantees of convergence than Perseus. While HSVI and SARSOP have been shown to outperform Perseus on certain problems, for the present problem we view the restriction to a single initial belief as a limitation. We will test SARSOP on the olfactory search problem in Ref.~\cite{loisy2023}.

Finally, we note that it may be possible to improve the performance of Perseus by projecting the beliefs onto a space of smaller dimension, using a form of non-negative matrix factorization (see, for example, \cite{li2007}).

\subsection{Other heuristic strategies}
In addition to those detailed in Sec.~\ref{sec:heuristic}, there are a litany of other heuristic strategies which we could have considered. Among these, we  tested the \emph{most likely state} policy which strikes towards the location with maximum belief,
\be
\pi_{\rm MLS}(b) = \pi^*_{\rm MDP}(\argmax_{s \in S} b(s)) ,
\ee
and \emph{action voting}, which selects the action most likely to be optimal according to the underlying MDP,
\be
\pi_{\rm AV}(b) = \argmax_{a \in A} \sum_{s \in S} b(s) \delta(\pi^*_{\rm MDP}(s), a),
\ee
where $\delta(\cdot,\cdot)$ is a Kronecker delta.

We found these policies to be too myopic/greedy and generally inferior to the others when applied to the search problem, so we did not present results for them in this work.

\subsection{Testing policies}\label{sec:testing}
In determining whether or not a policy is good, we must first specify the problem space we are interested in solving: namely, where the agent starts its search. We limit ourselves to problems that are neither unreasonably hard nor unreasonably easy in the following way: the agent always begins its search somewhere between two isocurves of detection likelihood $\ell={\rm Pr}(o|\mathbf{s}).$ Specifically, we (somewhat arbitrarily) select the range $0.006 \bar{S} < \ell < 0.02 \bar{S}.$ where, as a reminder, $\bar{S}$ is the nondimensionalized emission rate of the source (see Fig.~\ref{fig:ellipses}). The scaling with $\bar{S}$ ensures that the curves are approximately invariant when the emission rate is changed. For consistency, this condition is imposed on the agent's starting location both when we are testing the policy and when we are collecting beliefs in the initial phase of Perseus.
During training, at each iteration of Perseus \ref{eq:perseusValueIteration}, we evaluate the resulting policy in a couple of different ways. First, we randomly select (without replacement) an ensemble of 100 points lying within the acceptable isocurves, and trial the policy starting from each point ten times. (For each $S$, the set of starting points is held constant across policies and iterations for consistency.) This leads to 1000 Monte Carlo trials whose arrival times and rewards are then averaged, to get a sense of the overall performance of the policy. 

Because the ensemble averaging is poorly controlled, we also evaluate the policies on a small set of four fixed starting points at different distances from the source, using $1000$ trials each. For a more detailed view of the statistics, for a few select policies we extend the number of trials to $10000,$ which we use to construct the arrival time pdfs.

It should be pointed out that this prescription for evaluating performance is \emph{ad hoc} and in particular diverges somewhat from the rigorous definition of optimality for a POMDP. Formally, the optimal policy is optimal when conditioned on the prior, so the source should be drawn from the initial belief for a rigorous evaluation of optimality. We take this approach in Appendix~\ref{app:init}, which loosely follows the initialization used in \cite{loisy2021}. The careful reader should understand the approach taken in the main body of the present work as the specification of an interesting problem based on phenomenology, and the use of a POMDP solver as a heuristic to find an empirically best solution to this problem.

\section{Convergence of Perseus}\label{app:convergence}
In this section we show how a few key features of the Perseus policies evolve from iteration to iteration, while comparing different choices of the shaping function. These features include the performance on test problems and the Bellman error. We also compare performance for different choices of $\gamma$.

\subsection{Mean arrival times for an ensemble of starting points}
\FloatBarrier
In Fig.~{\ref{fig:low_ensemble_arrival} we show the evolution, over Perseus iterations, of the mean arrival time for an ensemble of 100 randomly-selected starting points. The agent searches from each starting point 10 times, yielding 10 estimates for the mean arrival time; the standard error over these 10 estimates is used as our error bar. Note that, to limit computation time, the agent's search was limited to 1000 timesteps.

For each emission rate, we show as a baseline the corresponding results for the heuristic which performed best at that emission rate. These are, respectively, Thompson sampling with $\tau=100$, space-aware infotaxis, and QMDP. We also tested Thompson sampling with $\tau=1$ and $\tau=10$; increasing $\tau$ beyond 100 had essentially no effect since the agent will almost never have to travel further than 100 units to reach a sampled point.

An immediate takeaway is that, with a good choice of reward shaping, significantly fewer iterations are required to achieve good performance on the search problem, relative to the unshaped baseline. In fact, on the relatively large grid studied here, we have found that no number of iterations seem to suffice for unshaped Perseus to ``catch up'' with Perseus using a good shaping.

Another observation is that introducing a reward shaping function can apparently reduce the stability of the policy from iteration to iteration, as evidenced by significant fluctuations in policy performance which were occasionally observed (the logarithmic shaping in the $\bar{S} = 25$ environment provides an extreme example of this). This behavior is more evident when $\gamma$ is larger and is usually intensified when the shaping is increased in magnitude. This, along with the concomitant need for additional hyperparameter tuning, appears to be the main drawback of using reward shaping.
\begin{figure}
    \centering
    \includegraphics[width=0.8\linewidth]{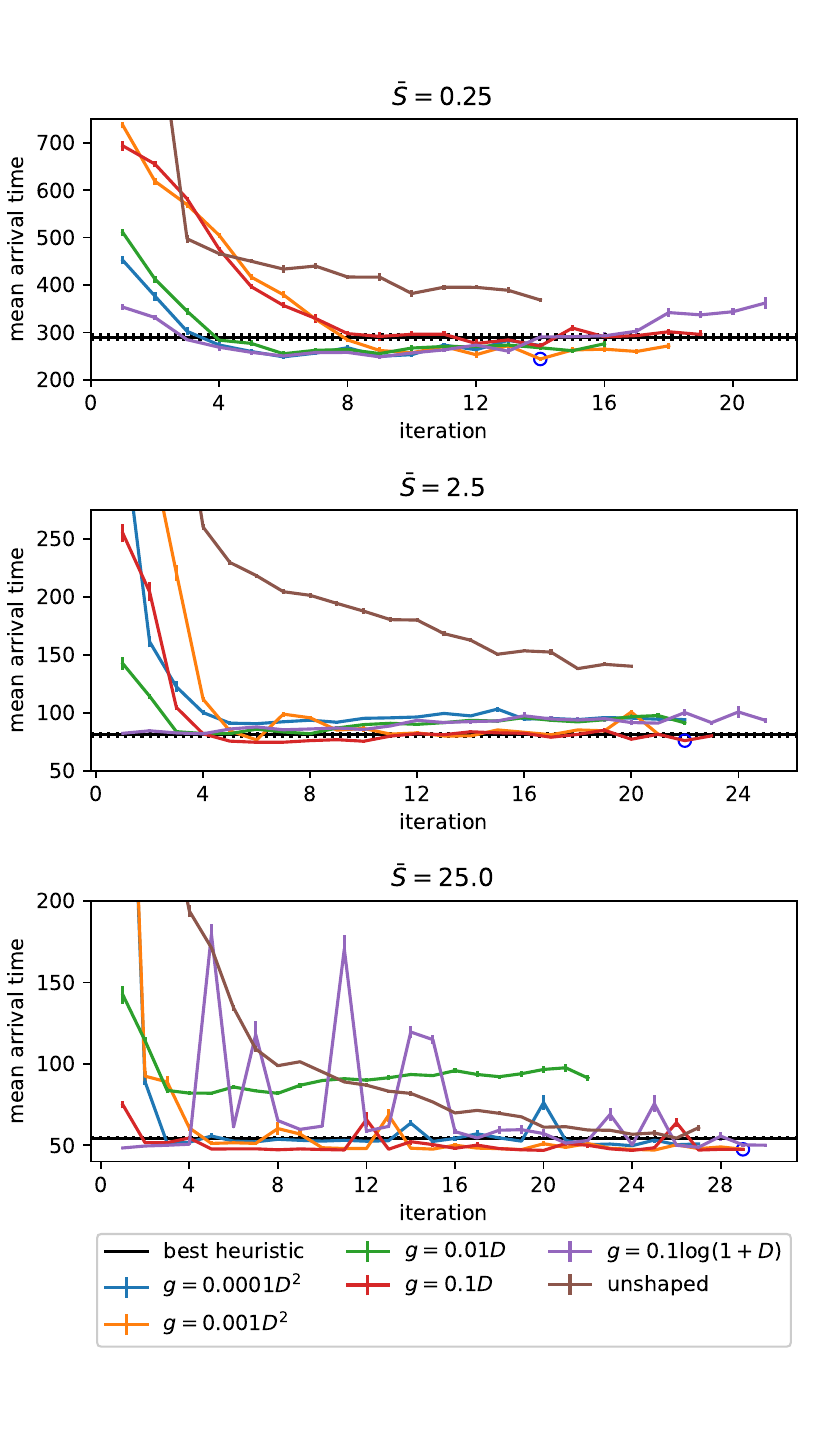}
    \caption{Mean arrival times from ensemble of starting points in the each environment, for several choices of reward shaping. Error bars represent the standard error of the mean as estimated by the variance across the ten trials. The policy chosen for testing is circled.}
    \label{fig:low_ensemble_arrival}
\end{figure}
\subsection{Bellman error}
To claim that the policies obtained using Perseus are near-optimal, one should confirm that the Bellman error (Eq.~\ref{eq:bellmanerror}) is decreasing from iteration to iteration. In Fig.~\ref{fig:training_bellman}, we show, for $\bar S = 2.5,$ the rms Bellman error on the belief set ${\cal B}$ and on beliefs encountered during testing. Borrowing from the lexicon of machine learning, we call these ``training'' and ``validation'' errors. 

\begin{figure}
    \centering
    \includegraphics[width=\linewidth]{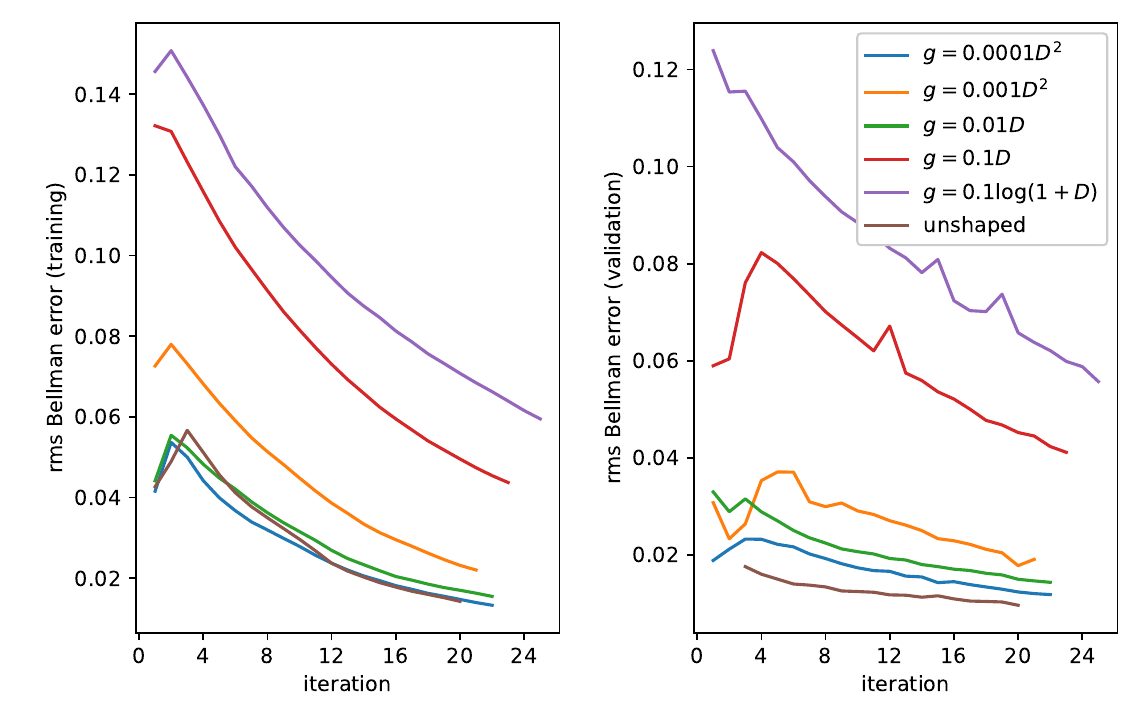}
    \caption{Rms Bellman training and validation errors, i.e.\ errors on ${\cal B}$ and those encountered during testing, respectively, for $\bar S = 2.5.$}
    \label{fig:training_bellman}
\end{figure}


\subsection{Dependence on $\gamma$}
On the left side of Fig.~\ref{fig:gammas_5.0} we show how the performance of Perseus, in the absence of a reward shaping function, depends on $\gamma$, for $\bar{S} = 2.5$. We plot the mean arrival times for the ensemble of starting points as a function of iteration for a few $\gamma,$ corresponding to a range of horizons 12.5--200. Note that performance is comparable for most of the $\gamma$'s, but $\gamma=0.99$ is too large and leads to poor stability/convergence. 

In the right side of the same figure, we show the same, but with a reward shaping $g=0.1 D$ turned on. While an excellent choice when $\gamma=0.98$, this shaping function leads to poor performance on all other $\gamma$'s. Thus it is clear that the best choice of shaping depends strongly on the choice of $\gamma$; one should select a $\gamma$ first.
\begin{figure}
    \centering
    \includegraphics[width=\linewidth]{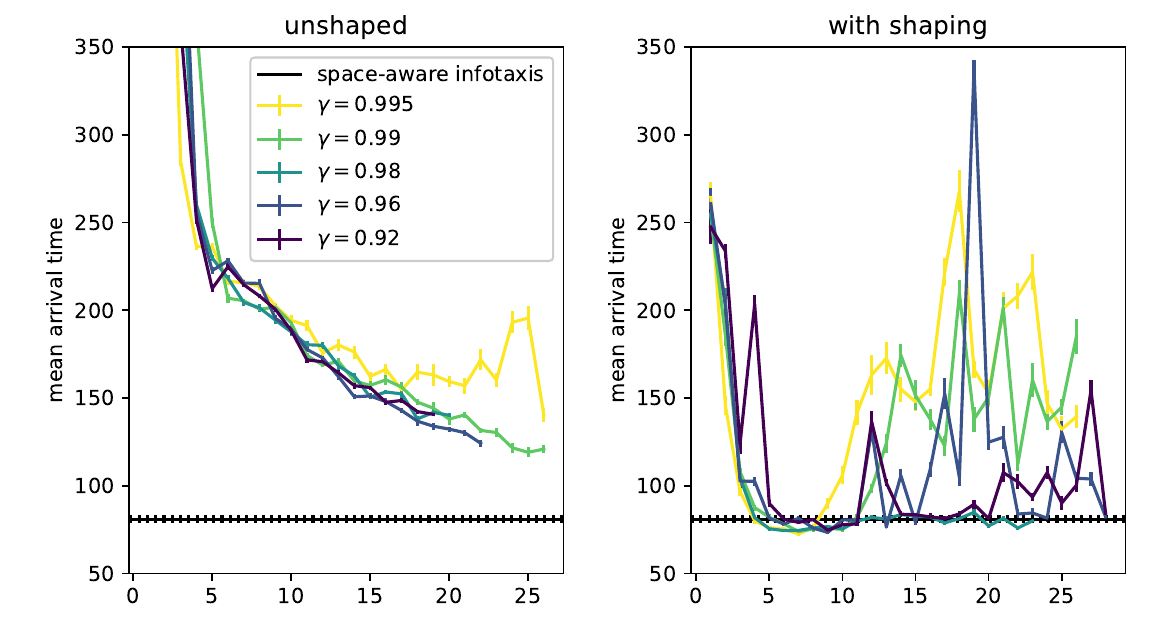}
    \caption{Evolution of mean arrival time from ensemble of starting points using Perseus, for $\bar S = 2.5$, and for several choices of $\gamma$. On the left, we do not use a reward shaping function, and on the right we use $g=0.1 D.$}
    \label{fig:gammas_5.0}
\end{figure}

\section{Alternate initialization}\label{app:init}
Previously in this work, we have explored an initialization of the belief where the agent waits in place until it receives a detection. Another initialization, as in Ref.~\cite{loisy2021}, is to force a detection at time zero. Under such a prescription, the initial belief will always be the same, and it will simply be proportional to the likelihood of detection. In this appendix, we briefly present performance results for the various policies using this approach. 

In Tables \ref{tab:alt_init_dilute}--\ref{tab:alt_init_high}, we show several properties of the arrival time distributions. To wit, we show the mean and several properties of the tails: the time $T_{90}$ corresponding to the 90th percentile of the pdf, the probability that the arrival time exceeds twice the mean obtained under Perseus, and the failure rate $\mathrm{Pr}(T\ge T_{\max})$ for $T_{\max}=2500.$ To be clear, we have followed Ref.~\cite{loisy2021} and, for each Monte Carlo trial, drawn the location of the source randomly from the initial belief, which is a more precise test of optimality.

The results are qualitatively very different from those obtained with the initialization we used in the main text. Across all three environments, one of the versions of infotaxis yielded the best results among heuristics, and Perseus performed equally well or slightly worse. In the dilute $\bar{S}=0.25$ environment, infotaxis and Perseus are nearly indistinguishable in terms of their performance --- infotaxis has a slightly lower mean arrival time but a slightly fatter tail. We find that SAI, while having a low mean arrival time when the agent does not fail, nevertheless suffers from a relatively high failure rate (more than 4\%) in this environment and thus is (arguably) not very competitive. QMDP fails more than 85\% of the time. Thompson sampling, the best heuristic in the corresponding environment in the main text, is less performant with this initialization. It is not terribly surprising that the performance Thompson sampling might depend on the prior, since the strategy depends on sampling from this prior. 

In the $\bar{S}=2.5$ environment, the performance of SAI and Perseus are now nearly indistinguishable. Infotaxis also performs well, and the other heuristic policies are not competitive.

Finally, in the $\bar{S}=25$ environment, SAI was slightly better than Perseus, and infotaxis lagged somewhat further behind. QMDP, which was the best heuristic for the high emission problem under the initialization in the main text, still suffers from a rather high failure rate.

These results illustrate the fact that the performance of various policies on the search problem depends strongly on the choice of prior, so may be interesting to study which prior comports best with observed insect behavior. It is remarkable how well infotaxis and SAI perform in this setting, and the results here strongly suggest that, depending on the emission rate, one or both of these heuristics are very close to optimal under this prior.

Other solvers, such as SARSOP, may be better suited for this initialization than Perseus due to there being only a single initial belief; this will be checked in \cite{loisy2023}.

\begin{table}
    \centering
    \begin{tabular}{|c|c|c|c|c|}
    \hline
         policy & $\mathbb{E}[T|T<T_{\max}]$ & $T_{90}$ & $\mathrm{Pr}(T>2 \langle T \rangle_{\rm perseus})$ & $\mathrm{Pr}(T\ge T_{\max})$  \\
         \hline
        MDP optimum & $ 21.6 \pm 0.1$ & - & - & - \\
       \textbf{Perseus} & $ \mathbf{203.2 \pm 1.8 }$& \textbf{539} & \textbf{0.162} & \textbf{0.0011} \\
        \textbf{infotaxis} & $ \mathbf{197.6 \pm 1.8} $ & \textbf{567} &\textbf{0.165} & $\mathbf{<5 \times 10^{-5}}$ \\
        SAI & $136.8 \pm 1.2 $ & 469 & 0.127 & 0.043 \\
        QMDP & $22.2 \pm 5.8 $ & 2500 & 0.858 & 0.857 \\
        Thompson, $\tau=1$ & $483.8 \pm 3.9$ & 1837 & 0.455 & 0.065 \\
        Thompson, $\tau=10$ & $347.0 \pm 2.8$ & 884 & 0.299 & 0.0086 \\
        Thompson, $\tau=100$ & $351.3 \pm 2.5$ & 795 & 0.311 & 0.0019 \\
        \hline
    \end{tabular}
    \caption{Performance statistics of various polices with the alternate initialization described in this appendix, with $\bar{S} = 0.25$. Note that $T_{\max}$ was set to 2500. The MDP optimum (i.e., the mean Manhattan distance) is shown for comparison.}
    \label{tab:alt_init_dilute}
\end{table}

\begin{table}
    \centering
    \begin{tabular}{|c|c|c|c|c|}
    \hline
         policy & $\mathbb{E}[T|T<T_{\max}]$ & $T_{90}$ & $\mathrm{Pr}(T>2 \langle T \rangle_{\rm perseus})$ & $\mathrm{Pr}(T\ge T_{\max})$  \\
         \hline
        MDP optimum & $23.3 \pm 0.1$ & - & - & - \\
        \textbf{Perseus} & $\mathbf{64.4 \pm 0.4}$ & \textbf{143} & \textbf{0.131} & $\mathbf{3 \times 10^{-4}}$ \\
        \textbf{infotaxis} & $\mathbf{66.8 \pm 0.5}$ & \textbf{155} & \textbf{0.163} & $\mathbf{<5 \times 10^{-5}}$ \\
        \textbf{SAI} & $\mathbf{62.1 \pm 0.4 }$ & \textbf{145} & \textbf{0.139} & $\mathbf{1.5 \times 10^{-4}}$ \\
        QMDP & $86.7 \pm 0.8 $ & 2500 & 0.444 & 0.281 \\
        Thompson, $\tau=1$ & $138.9 \pm 1.4$ & 301 & 0.359 & 0.0027 \\
        Thompson, $\tau=10$ & $102.0 \pm 0.7$ & 212 & 0.284 & $<5 \times 10^{-5}$ \\
        Thompson, $\tau=100$ & $124.3 \pm 0.7$ & 241 & 0.396 & $<5 \times 10^{-5}$ \\
        \hline
    \end{tabular}
    \caption{Same as Table \ref{tab:alt_init_dilute}, but with $\bar{S} = 2.5$. }
    \label{tab:alt_init_med}
\end{table}

\begin{table}
    \centering
    \begin{tabular}{|c|c|c|c|c|}
    \hline
        policy & $\mathbb{E}[T|T<T_{\max}]$ & $T_{90}$ & $\mathrm{Pr}(T>2 \langle T \rangle_{\rm perseus})$ & $\mathrm{Pr}(T\ge T_{\max})$  \\ 
        \hline
        MDP optimum & $29.8 \pm 0.1$ & - & - & - \\ 
        \textbf{Perseus} & $\mathbf{43.9 \pm 0.2}$ & \textbf{76} & \textbf{0.070} & $\mathbf{4.5 \times 10^{-4}}$ \\
        infotaxis & $48.5 \pm 0.2$ & 92 & 0.125 & 0.0018 \\
        \textbf{SAI} & $\mathbf{41.6 \pm 0.2 }$ & \textbf{76} & \textbf{0.044} & $\mathbf{<5 \times 10^{-5}}$ \\
        QMDP & $51.2 \pm 0.3 $ & 100 & 0.138 & 0.031 \\
        Thompson, $\tau=1$ & $70.9 \pm 0.5$ & 122 & 0.281 & $5 \times 10^{-5}$ \\
        Thompson, $\tau=10$ & $58.07 \pm 0.2$ & 99 & 0.171 & $<5 \times 10^{-5}$ \\
        Thompson, $\tau=100$ & $82.7 \pm 0.3$ & 136 & 0.398 & $<5 \times 10^{-5}$ \\
        \hline
    \end{tabular}
    \caption{Same as Table \ref{tab:alt_init_dilute}, but with $\bar{S} = 25$.}
    \label{tab:alt_init_high}
\end{table}

\acknowledgements{RAH gratefully acknowledges useful discussions with Aurore Loisy, Fabio Bonaccorso, Michele Buzzicotti, and Antonio Costa. RAH and LB received funding from the European Research Council (ERC) under the European Union's Horizon 2020 research and innovation programme (Grant Agreement No.\ 882340). AC acknowledges funding from the European Union’s Horizon 2020 research and innovation programme under the Marie Skłodowska-Curie grant agreement No.\ 956457.}

\end{document}